\documentclass[onecolumn,showpacs,preprintnumbers]{revtex4}
\usepackage{amsfonts}
\usepackage{amsmath}
\usepackage{graphicx}
\usepackage{dcolumn}
\usepackage{bm}
\usepackage{mathrsfs}
\usepackage{epsfig}

\begin{document}

\preprint{APS/123-QED}
\title{Kinetic description of quasi-stationary axisymmetric collisionless \\
accretion disk plasmas with arbitrary magnetic field configurations}
\author{Claudio Cremaschini}
\altaffiliation[Also at ]{Consortium for Magnetofluid Dynamics, University of Trieste, Italy}
\affiliation{International School for Advanced Studies (SISSA), Trieste, Italy}
\author{John C. Miller}
\altaffiliation[Also at ]{Consortium for Magnetofluid Dynamics, University of Trieste, Italy}
\affiliation{International School for Advanced Studies (SISSA) and INFN, Trieste, Italy \\
Department of Physics (Astrophysics), University of Oxford, U.K.}
\author{Massimo Tessarotto}
\altaffiliation[Also at ]{Consortium for Magnetofluid Dynamics, University of Trieste, Italy}
\affiliation{Department of Mathematics and Informatics, University of Trieste, Italy}
\date{\today }

\begin{abstract}
A kinetic treatment is developed for collisionless magnetized plasmas
occurring in high-temperature, low-density astrophysical accretion disks,
such as are thought to be present in some radiatively-inefficient accretion
flows onto black holes. Quasi-stationary configurations are investigated,
within the framework of a Vlasov-Maxwell description. The plasma is taken to
be axisymmetric and subject to the action of slowly time-varying
gravitational and electromagnetic fields. The magnetic field is assumed to
be characterized by a family of locally nested but open magnetic surfaces.
The slow collisionless dynamics of these plasmas is investigated, yielding a
reduced gyrokinetic Vlasov equation for the kinetic distribution function.
For doing this, an asymptotic quasi-stationary solution is first determined,
represented by a generalized bi-Maxwellian distribution expressed in terms
of the relevant adiabatic invariants. The existence of the solution is shown
to depend on having suitable kinetic constraints and conditions leading to
particle trapping phenomena. With this solution one can treat temperature
anisotropy, toroidal and poloidal flow velocities and finite Larmor-radius
effects. An asymptotic expansion for the distribution function permits
analytic evaluation of all of the relevant fluid fields. Basic theoretical
features of the solution and their astrophysical implications are discussed.
As an application, the possibility of describing the dynamics of slowly
time-varying accretion flows and the self-generation of magnetic field by
means of a \textquotedblleft kinetic dynamo effect\textquotedblright\ is
discussed. Both effects are shown to be related to intrinsically-kinetic
physical mechanisms.
\end{abstract}

\maketitle










\section{Introduction}

This paper is part of an investigation concerning the theoretical
formulation of kinetic theory for collisionless astrophysical plasmas in
accretion disks (ADs) around compact objects, and its application to the
study of their equilibrium properties and dynamical evolution. Note that
what is meant here by the word \textquotedblleft
equilibrium\textquotedblright\ is in general a stationary-flow solution,
which can also include a stationary radial accretion velocity.

In contrast with the majority of previous treatments, which are based on
fluid approaches within the context of hydrodynamics (HD) or
magnetohydrodynamics (MHD) \cite{Frank,Vietri,Miller2001,Naso,Cremasch2008-1}%
, here we adopt a kinetic approach. This provides a phase-space treatment
allowing us to formulate a consistent description of plasma dynamics.
Kinetic theory is essential for studying both stationary configurations and
dynamical evolution of plasmas when kinetic effects are relevant, such as
ones associated with conservation of particle adiabatic invariants,
temperature anisotropy, finite Larmor-radius (FLR) effects (as pointed out
in Ref.\cite{Cremasch2009}) and kinetic trapping phenomena. These properties
are relevant for magnetized plasmas and in particular for those arising in
ADs \cite{Quataert2002,Quataert2007,Quataert2007B} whenever the plasma is
regarded as collisionless or weakly collisional \cite%
{Cremasch2008-2,Cremasch2009}.

In the context of astrophysical ADs, there are several examples of
collisionless plasmas of this kind, with both strong and weak magnetic
fields. One is the case of radiatively inefficient accretion flows (RIAFs)
\cite{NarayanA,NarayanB}, in geometrically thick disks around black holes
consisting of two-temperature plasma, with the ion temperature being much
higher than the electron one, and the timescale of the Coulomb collision
frequency being much longer than the inflow time. Other interesting
applications occur in ADs around neutron stars and white dwarfs: in the
inner regions of such disks, where the magnetic field of the central object
becomes dominant, ions and electrons can be collisionally decoupled and
sustain different temperatures. This happens, in particular, if the
radiative cooling time-scale of the electrons is much shorter than the
time-scale for electron-ion collisions. In this way, electrons and ions are
thermally decoupled: the two charged species acquire unequal temperatures
and the accretion flow becomes a two-temperature flow \cite%
{Saxton2005,Saxton2007}.

In our earlier paper (Ref.\cite{Cremasch2009}, hereafter referred to as
Paper I), we presented preliminary results in this direction, concerning
formulation of kinetic theory for investigating stationary solutions for
collisionless AD plasmas, focusing on configurations with locally-closed
magnetic flux surfaces. The present paper is intended as a continuation of
the previous one, with the aim of generalizing the previous solution to
arbitrary magnetic field configurations, which are no longer restricted to
localized spatial domains in the disk. We refer to Fig.1 below and the
discussion in Section II for an explicit comparison of the two
configurations. More specifically, in Paper I the treatment concerned
collisionless magnetized plasmas characterized by locally closed and nested
magnetic surfaces. For such configurations, it was shown that suitable
kinetic distribution functions (KDFs) are permitted, describing both kinetic
and gyrokinetic (GK) equilibria (see definition in Paper I), which are
represented by generalized Maxwellian and bi-Maxwellian KDFs. A main feature
was the inclusion in the kinetic treatment of both temperature anisotropy
and FLR effects. In particular, in Paper I and in Ref.\cite{Catania1}, it
was proved that these equilibria can sustain a stationary kinetic dynamo. As
a basic consequence, it was found that both toroidal and poloidal
equilibrium magnetic fields can be self generated for quasi-neutral plasmas,
\textit{without ongoing instabilities and/or turbulence phenomena}. In
particular, in closed nested field configurations [and hence in the local
absence of net accretion flow], the toroidal field component was found to be
produced by diamagnetic effects driven by the species temperature
anisotropies. As a further development, in Paper I and in Ref.\cite{Catania2}%
, it was pointed out that the kinetic treatment allows one to construct
exact fluid equilibria (identically satisfying the corresponding fluid
equations). Using a perturbative expansion, a well-defined set of kinetic
closure conditions was determined analytically for the relevant stationary
moment equations.

\bigskip

\subsection{Accretion disks in astrophysics}

Despite more than forty years of observations and theoretical
investigations, there is a lot remaining to be understood about the physical
processes governing the structure and evolution of ADs. They are observed in
a wide range of astrophysical contexts \cite{Frank} and consist of plasma
orbiting a central object with the velocities of the inward accretion flow
usually being much smaller than the rotational velocities. In order for the
accretion to happen, there needs to be a net outward transport of angular
momentum and there are several conceivable mechanisms for producing this
(see for example \cite{Rebusco09,Coppi09}). The most obvious one is fluid
viscosity, but this would need to be an \textquotedblleft
anomalous\textquotedblright viscosity, driven by some type of turbulence,
rather than a standard viscosity connected with Coulomb collisions (Spitzer
viscosity) which would be much too small to explain the observed accretion
rates under the conditions actually found in accretion disks (see \cite%
{Lomi09} for a review of turbulence mechanisms). However, other
collisionless physical mechanisms are possible in principle, such as kinetic
instabilities, radiation effects and magnetic reconnection. The aim of the
present paper is to help in preparing the way for a discussion of these. We
focus on AD plasmas immersed in slowly time-varying magnetic fields,
characterized locally by open nested magnetic surfaces. For these systems,
no kinetic treatment has been available up to now. The origin of their
magnetic fields varies depending on the type of the central object: in the
case of black holes, the fields are only ones self-generated by currents in
the plasma itself via dynamo effects, while with neutron stars and white
dwarfs there can also be a magnetic field intrinsic to the central object
\cite{Frank,Vietri}. The interplay between magnetic fields and accretion
plasmas can affect the overall velocity profile of the disk, as well as
giving rise to species-dependent velocities and rotational frequencies \cite%
{Naso,Cremasch2008-1, Cremasch2008-2}. Moreover, the magnetic field can be a
source of anisotropies in the KDF and allow particular symmetries which
influence both the single particle and collective plasma behavior. The
transport of angular momentum, the accretion flow and the possible
generation of jets \cite{Pelletier01,Pelletier02,Pelletier03} are all
strongly dependent on the magnetic field structure and so magnetic fields
play an important role for AD physics.

\bigskip

\subsection{Goals and scheme of the presentation}

The purpose of this paper is to formulate a comprehensive kinetic treatment
for collisionless axisymmetric AD plasmas including both accretion flows and
collisionless dynamo effects. We include general relative orderings between
the magnitudes of the external and self-generated magnetic fields and allow
the magnetic field be non-uniform and slowly time-varying while possessing
locally nested open magnetic surfaces.

Extending the investigation developed in Paper I, we do this by constructing
particular quasi-stationary solutions of the Vlasov-Maxwell equations,
characterized by generalized bi-Maxwellian phase-space distributions (see
also Paper I), which are referred to here as quasi-stationary asymptotic
KDFs (QSA-KDFs). As discussed below, the functional form of these solutions
is physically motivated. We will show that this makes possible the explicit
inclusion of both \textit{temperature anisotropies} and \textit{parallel
velocity perturbations} in the QSA-KDFs (see the definition below in Section
4). This is done, first, by developing an \textquotedblleft ad
hoc\textquotedblright\ formulation for GK theory in the presence of a
gravitational field, making it possible to directly construct the relevant
particle guiding-center adiabatic invariants. The QSA-KDFs are then
expressed in terms of these. Remarkably, this allows also the consistent
treatment of trapping phenomena due to spatial variations both of the
magnetic field and of the total effective potential (gravitational EM
trapping). Second, the QSA-KDFs are constructed by imposing appropriate
\textit{kinetic constraints} (see Section 4), requiring that suitable
\textit{structure functions} (see below) which enter the definition of the
QSA-KDFs, depend only on the azimuthal canonical momentum and total particle
energy. By invoking suitable perturbative expansions, it follows that the
relevant moments and moment equations can be evaluated analytically. The
solution thus obtained can be used for investigating the quasi-stationary
dynamics of magnetized AD plasmas, including description of quasi-stationary
accretion flows and \textquotedblleft {kinetic dynamo
effects\textquotedblright }\ allowing for the generation of finite poloidal
and toroidal magnetic fields. In particular, the kinetic theory predicts the
possibility of pure matter inflows as well as the independent coexistence of
both inflows and outflows.

The paper is organized as follows. In Section 2 we summarize the basic
assumptions and definitions of the theory. In Section 3 we formulate the GK
theory for magnetized accretion disk plasmas, deriving the relevant
integrals of motion and guiding-center adiabatic invariants and discussing
the particle trapping phenomenon. Section 4 deals with the construction of a
generalized asymptotic stationary KDF, with the inclusion of parallel
velocity perturbations and the adoption of suitable kinetic constraints. In
Section 5 we give an analytic expansion for the KDF and discuss its main
features. Section 6 deals with the relationship between kinetic theory and
the corresponding fluid treatment, which concerns the validity of moment
equations and the analytic calculation of fluid fields. Section 7 is
dedicated to discussing the temporal evolution of the GK equilibria and
derivation of the dynamical equation for the GK KDF. In Section 8 we
investigate the implication of the kinetic solution for the Ampere equation
and the existence of the kinetic dynamo effect. Then, in Section 9 we
discuss the treatment of quasi-stationary accretion flow within the
present formulation, showing that solutions with net radial accretion are
admitted consistently with the constraints imposed by the Maxwell equations.
Finally, Section 10 contains a summary of the main results with closing
remarks.

\bigskip

\section{Basic assumptions and definitions}

Ignoring possible weakly-dissipative effects (Coulomb collisions and
turbulence), we shall assume that the KDF and the EM fields associated with
the plasma obey the system of Vlasov-Maxwell equations, with Maxwell's
equations being considered in the quasi-static approximation. For
definiteness, we shall consider here a plasma consisting of at least two
species of charged particles: one species of ions ($i$) and one of electrons
($e$).

Following the treatment presented in Paper I, we shall take the AD plasma to
be: a) \emph{non-relativistic}, in the sense that it has non--relativistic
species flow velocities, that the gravitational field can be treated within
the classical Newtonian theory, and that the non-relativistic Vlasov kinetic
equation is used as the dynamical equation for the KDF;\ b) \emph{%
collisionless}, so that the mean free path of the plasma particles is much
longer than the largest characteristic scale length of the plasma; c)\ \emph{%
axisymmetric}, so that the relevant dynamical variables characterizing the
plasma (e.g., the fluid fields) are independent of the azimuthal angle $%
\varphi ,$ when referred to a set of cylindrical coordinates $(R,\varphi ,z)$%
; d) acted on by both gravitational and EM fields.

The kinetic formulation is intrinsically asymptotic. This means that the
theory (in particular the GK theory formulated in the next section) is
characterized by a suitable species-dependent dimensionless physical
parameter $\varepsilon _{M,s}\equiv \frac{r_{Ls}}{L}\ll 1 $, where $s=i,e$
denotes the species index. Here $r_{Ls}=v_{\perp ths}/\Omega _{cs}$ is the
species average Larmor radius, with $v_{\perp ths}=\left\{ T_{\perp
s}/M_{s}\right\} ^{1/2}$ denoting the species thermal velocity perpendicular
to the magnetic field and $\Omega _{cs}=Z_{s}eB/M_{s}c$ being the species
Larmor frequency. Moreover, $L $ is the characteristic length-scale of the
spatial inhomogeneities of the EM field, defined as $L\sim L_{B}\sim L_{E}$,
where $L_{B}$ and $L_{E}$ are the characteristic magnitudes of the gradients
of the absolute values of the magnetic field $\mathbf{B}\left( \mathbf{x}%
,t\right) $ and the electric field $\mathbf{E}\left( \mathbf{x},t\right) $,
defined as $\frac{1}{L_{B}} \equiv \max \left\{ \left\vert \frac{\partial }{%
\partial r_{i}}\ln B\right\vert ,i=1,3\right\} $ and $\frac{1}{L_{E}}\equiv
\max \left\{ \left\vert \frac{\partial }{\partial r_{i}}\ln E\right\vert
,i=1,3\right\} $, where the vector $\mathbf{x}$ denotes $\mathbf{x}=\left(
R,z\right) $. Then, in analogy with Paper I, we define a unique parameter $%
\varepsilon _{M}\equiv \max \left\{ \varepsilon _{M,s},s=i,e\right\} $. For
temperatures and magnetic fields typical of AD plasmas, we have $%
0<\varepsilon _{M}\ll 1$.

In the following we will focus on solutions for the equilibrium magnetic
field $\mathbf{B}$ which admit, at least locally, a family of nested and
\textit{open }axisymmetric toroidal magnetic surfaces $\left\{ \psi (\mathbf{%
\ x})\right\} \equiv \left\{ \psi (\mathbf{x})=const.\right\} $, where $\psi
$ denotes the poloidal magnetic flux of $\mathbf{B}$. See Fig. 1 for a
schematic comparison between the configuration of locally closed magnetic
surfaces considered in Paper I and the case of open magnetic surfaces
analyzed in the present study. A set of magnetic coordinates ($\psi ,\varphi
,\vartheta $) can be defined locally, where $\vartheta $ is a curvilinear
angle-like coordinate on the magnetic surfaces $\psi (\mathbf{x})=const.$
Each relevant physical quantity $G(\mathbf{x},t)$ can then be conveniently
expressed either in terms of the cylindrical coordinates or as a function of
the magnetic coordinates, i.e. $G(\mathbf{x},t)=\overline{G}\left( \psi
,\vartheta ,t\right) ,$ where the $\varphi $ dependence has been suppressed
due to the axisymmetry.

\begin{figure}[tbp]
\centering
\includegraphics[width=3.5in,height=2.5in]{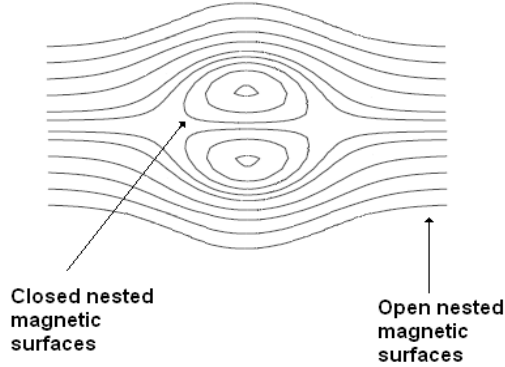} 
\caption{Schematic comparison between the configuration of locally closed
magnetic surfaces considered in Paper I and the case of open magnetic
surfaces analysed in the present study.}
\end{figure}

We require the EM field to be slowly varying in time, i.e., to be of the
form
\begin{equation}
\left[ \mathbf{E}(\mathbf{x},\varepsilon _{M}^{k}t),\mathbf{B}(\mathbf{x}%
,\varepsilon _{M}^{k}t)\right] ,  \label{b0}
\end{equation}%
with $k\geq 1$ being a suitable integer. This time dependence is connected
with either external sources or boundary conditions for the KDF. In
particular, we shall assume that the magnetic field is of the form
\begin{equation}
\left. \mathbf{B}\equiv \nabla \times \mathbf{A}=\mathbf{B}^{self}(\mathbf{x}%
,\varepsilon _{M}^{k}t)+\mathbf{B}^{ext}(\mathbf{x},\varepsilon
_{M}^{k}t),\right.  \label{b1}
\end{equation}%
where $\mathbf{B}^{self}$ and $\mathbf{B}^{ext}$ denote the self-generated
magnetic field produced by the AD plasma and a finite external magnetic
field produced by the central object (in the case of neutron stars or white
dwarfs). For greater generality, we shall not prescribe any relative
orderings between the various components of the total magnetic field, which
are taken to be of the form%
\begin{eqnarray}
&&\left. \mathbf{B}^{self}=I(\mathbf{x},\varepsilon _{M}^{k}t)\nabla \varphi
+\nabla \psi _{p}(\mathbf{x},\varepsilon _{M}^{k}t)\times \nabla \varphi
,\right.  \label{bself} \\
&&\left. \mathbf{B}^{ext}=\nabla \psi _{D}(\mathbf{x},\varepsilon
_{M}^{k}t)\times \nabla \varphi .\right.
\end{eqnarray}%
In particular, here $\mathbf{B}_{T}\equiv I(\mathbf{x},\varepsilon
_{M}^{k}t)\nabla \varphi $ and $\mathbf{B}_{P}\equiv \nabla \psi _{p}(%
\mathbf{x},\varepsilon _{M}^{k}t)\times \nabla \varphi $ are the toroidal
and poloidal components of the self-field, while the external magnetic field
$\mathbf{B}^{ext}$ has to be purely poloidal, as a consequence of the
axisymmetry, and is defined in terms of the vacuum potential $\psi _{D}(%
\mathbf{x},\varepsilon _{M}^{k}t)$. As a consequence, the magnetic field can
also be written in the equivalent form
\begin{equation}
\mathbf{B}=I(\mathbf{x},\varepsilon _{M}^{k}t)\nabla \varphi +\nabla \psi (%
\mathbf{x},\varepsilon _{M}^{k}t)\times \nabla \varphi ,  \label{B FIELD}
\end{equation}%
where the function $\psi (\mathbf{x},\varepsilon _{M}^{k}t)$ is defined as $%
\psi (\mathbf{x},\varepsilon _{M}^{k}t)\equiv \psi _{p}(\mathbf{x}%
,\varepsilon _{M}^{k}t)+\psi _{D}(\mathbf{x},\varepsilon _{M}^{k}t),$ with $%
k\geq 1$ and $(\psi ,\varphi ,\vartheta )$ defining a set of local magnetic
coordinates (as implied by the equation $\mathbf{B}\cdot \nabla \psi =0$
which is identically satisfied). Also, it is assumed that the charged
particles of the plasma are subject to the action of \emph{effective} \emph{%
EM potentials} $\left\{ \Phi _{s}^{eff}(\mathbf{x},\varepsilon _{M}^{k}t),%
\mathbf{A}(\mathbf{x},\varepsilon _{M}^{k}t)\right\} ,$ where $\mathbf{A}(%
\mathbf{x},\varepsilon _{M}^{k}t)$ is the vector potential corresponding to
the magnetic field of Eq.(\ref{B FIELD}), while $\Phi _{s}^{eff}(\mathbf{x}%
,\varepsilon _{M}^{k}t)$ is given by%
\begin{equation}
\Phi _{s}^{eff}(\mathbf{x},\varepsilon _{M}^{k}t)=\Phi (\mathbf{x}%
,\varepsilon _{M}^{k}t)+\frac{M_{s}}{Z_{s}e}\Phi _{G}(\mathbf{x},\varepsilon
_{M}^{k}t),  \label{efp}
\end{equation}%
with $\Phi _{s}^{eff}(\mathbf{x},\varepsilon _{M}^{k}t),$ $\Phi (\mathbf{x}%
,\varepsilon _{M}^{k}t)$ and $\Phi _{G}(\mathbf{x},\varepsilon _{M}^{k}t)$
denoting the \emph{effective} electrostatic potential and the electrostatic
and generalized gravitational potentials (the latter, in principle, being
produced both by the central object and the accretion disk). The effective
electric field $\mathbf{E}_{s}^{eff}$ can then be defined as%
\begin{equation}
\mathbf{E}_{s}^{eff}\equiv -\nabla \Phi _{s}^{eff}-\frac{1}{c}\frac{\partial
\mathbf{A}}{\partial t}.  \label{electric}
\end{equation}

\bigskip

\section{GK theory for magnetized accretion disk plasmas}

In this section we recall the GK theory appropriate for the description of
AD plasmas. Its formulation is in fact a prerequisite for the construction
of the kinetic quasi-stationary equilibria to be developed later. The
appropriate generalization of GK theory allowing for the presence of strong
gravitational fields should in principle be based on a covariant formulation
[see \cite{Beklemishev1999,Beklemishev2004, Cremaschini2006,Cremaschini2008}%
]. However, for non-relativistic plasmas within a gravitational field, the
appropriate formulation can also be directly recovered via a suitable
reformulation of the standard non-relativistic theory holding for
magnetically confined plasmas \cite%
{Catto1978,Littlejohn1979,Littlejohn1981,Littlejohn1983,Dubin1983,Hahm1988,Bernstein1985,Balescu,Meiss1990}%
.

In this case, the appropriate particle Lagrangian function can be
represented in terms of the effective EM potentials $\left\{ \Phi _{s}^{eff}(%
\mathbf{x},\varepsilon _{M}^{k}t),\mathbf{A}(\mathbf{x},\varepsilon
_{M}^{k}t)\right\} ,$ with $k\geq 1$, where $\Phi _{s}^{eff}$ is defined in
Eq.(\ref{efp}). In terms of the hybrid variables $\mathbf{z}\equiv ($\textbf{%
$\mathbf{x},$}$\mathbf{v})$ (with $\mathbf{x}$ and $\mathbf{v}$ denoting
respectively the particle position and velocity vectors), this is expressed
as%
\begin{equation}
\left. \mathcal{L}_{s}(\mathbf{z},\frac{d}{dt}\mathbf{z},\varepsilon
_{M}^{k}t)\equiv \mathbf{\dot{r}\cdot }\mathbf{P}_{s}-\mathcal{H}_{s}(%
\mathbf{z},\varepsilon _{M}^{k}t),\right.  \label{LAGRANGIAN}
\end{equation}%
where $\mathbf{P}_{s}\equiv \left[ M_{s}\mathbf{v}+\frac{Z_{s}e}{c}\mathbf{A}%
(\mathbf{x},\varepsilon _{M}^{k}t)\right] $ and%
\begin{equation}
\left. \mathcal{H}_{s}(\mathbf{z},\varepsilon _{M}^{k}t)=\frac{M_{s}}{2}v^{2}%
\mathbf{+}{Z_{s}e}\Phi _{s}^{eff}(\mathbf{x},\varepsilon _{M}^{k}t)\right.
\label{HA MILTONIAN}
\end{equation}%
denotes the corresponding Hamiltonian function in hybrid variables. The GK
treatment for the Lagrangian (\ref{LAGRANGIAN}) involves the construction -
in terms of a perturbative expansion determined by means of a power series
in $\varepsilon _{M}$ - of a diffeomorphism of the form
\begin{equation}
\mathbf{z}\equiv (\mathbf{\mathbf{r},v})\mathbf{\rightarrow z}^{\prime
}\equiv (\mathbf{r}^{\prime }\mathbf{,v}^{\prime })\mathbf{,}
\label{gyrokinetic transformation}
\end{equation}%
referred to as the \emph{GK transformation}. Note that, in the following, we
shall use a prime \textquotedblleft\ $^{\prime }$ \textquotedblright\ to
denote a dynamical variable defined at the \emph{guiding-center position} $%
\mathbf{r}^{\prime }$ (or $\mathbf{x}^{\prime }$ in axisymmetry). Here, by
definition, the transformed variables $\mathbf{z}^{\prime }$\textbf{\ }(%
\emph{GK state}) are constructed so that their time derivatives to the
relevant order in $\varepsilon _{M}$ have at least one ignorable coordinate
(a suitably-defined gyrophase $\phi ^{\prime }$). As an illustration, we
show the formulation of the perturbative theory to leading-order in $%
\varepsilon _{M}.$ In this case the GK transformation becomes simply%
\begin{equation}
\left\{
\begin{array}{c}
\mathbf{r}=\mathbf{r}^{\prime }-\frac{\mathbf{w}^{\prime }\times \mathbf{b}%
^{\prime }}{\Omega _{cs}^{\prime }}, \\
\mathbf{v}=u^{\prime }\mathbf{b}^{\prime }+\mathbf{w}^{\prime }+\mathbf{V}%
_{eff}^{\prime },%
\end{array}%
\right.  \label{trasformazionui guirocinetiche}
\end{equation}%
where $\mathbf{w}^{\prime }=w^{\prime }\cos \phi ^{\prime }\mathbf{e}%
_{1}^{\prime }+w^{\prime }\sin \phi ^{\prime }\mathbf{e}_{2}^{\prime }$,
with $\phi ^{\prime }$ denoting the gyrophase angle. In the following, the
GK transformation will be performed on all phase-space variables $\mathbf{z}%
\equiv (\mathbf{\mathbf{r},v})$, \textit{except} for the azimuthal angle $%
\varphi $ which is left unchanged \cite{Catto1987} and is therefore to be
considered as one of the GK variables. Here $\mathbf{b}^{\prime }=\mathbf{b}(%
\mathbf{x}^{\prime },\varepsilon _{M}^{k}t),$ with $\mathbf{b}(\mathbf{x}%
,\varepsilon _{M}^{k}t)\mathbf{\equiv B}($\textbf{$\mathbf{x},$}$\varepsilon
_{M}^{k}t)\mathbf{/}B(\mathbf{x},\varepsilon _{M}^{k}t),$ while $\Omega
_{cs}^{\prime }=\frac{Z_{s}eB^{\prime }}{M_{s}c}$ and $\mathbf{V}%
_{eff}^{\prime }$ are respectively the guiding-center Larmor frequency and
the \emph{effective drift velocity} produced by $\mathbf{E}_{s}^{^{\prime
}eff},$ namely%
\begin{equation}
\mathbf{V}_{eff}^{\prime }(\mathbf{x},\varepsilon _{M}^{k}t)\equiv \frac{c}{{%
B}^{\prime }}\mathbf{E}_{s}^{^{\prime }eff}\times \mathbf{b}^{\prime }.
\label{V^drift}
\end{equation}%
The rest of the notation is standard, with $u^{\prime }$ and $\mathbf{w}%
^{\prime }$ denoting respectively the parallel and perpendicular
(guiding-center) velocities, both defined relative to the frame locally
moving with velocity $\mathbf{V}_{eff}^{\prime }$. It follows that, when
expressed in terms of the GK variables $\mathbf{z}^{\prime }$, the GK
Lagrangian and Hamiltonian functions, $\mathcal{L}_{s}^{\prime }$ and $%
\mathcal{H}_{s}^{\prime }$, can be evaluated with the desired order of
accuracy. In particular, to leading-order, i.e. neglecting corrections of $%
O(\varepsilon _{M}^{n})$ with $n\geq 1$, $\mathcal{L}_{s}^{\prime }=\mathcal{%
L}_{s}^{\prime (1)}+O(\varepsilon _{M})$ and $\mathcal{H}_{s}^{\prime }=%
\mathcal{H}_{s}^{\prime (1)}+O(\varepsilon _{M})$, where $\mathcal{L}%
_{s}^{\prime (1)}$ and $\mathcal{H}_{s}^{\prime (1)}$ recover the customary
expressions%
\begin{equation}
\mathcal{L}_{s}^{\prime (1)}\equiv \overset{.}{\mathbf{r}}^{\prime }\mathbf{%
\cdot }\frac{Z_{s}e}{c}\mathbf{A}_{s}^{^{\prime }\ast }-\frac{{\overset{%
\cdot }{\phi ^{\prime }}}}{\Omega _{cs}^{\prime }}m_{s}^{\prime }B^{\prime }-%
\mathcal{H}_{s}^{\prime (1)},  \label{LAGR girocinetica}
\end{equation}%
\begin{equation}
\mathcal{H}_{s}^{\prime (1)}\equiv m_{s}^{\prime }B^{\prime }+\frac{M_{s}}{2}%
\left( u^{\prime }\mathbf{b}^{\prime }+\mathbf{V}_{eff}^{\prime }\right)
^{2}+{Z_{s}e}\Phi _{s}^{\prime \ast },  \label{HAM giricinetica}
\end{equation}%
with the magnetic moment $m_{s}^{\prime }\cong \mu _{s}^{\prime }\equiv
\frac{M_{s}w^{\prime 2}}{2B^{\prime }}$ to leading order, while the
gyrophase-independent \textit{modified EM potentials} $\left( \Phi
_{s}^{\prime \ast },\mathbf{A}_{s}^{^{\prime }\ast }\right) $ are%
\begin{eqnarray}
\Phi _{s}^{\prime \ast } &\cong &\Phi _{s}^{\prime eff}, \\
\mathbf{A}_{s}^{^{\prime }\ast } &\cong &\mathbf{A}^{\prime }+\frac{M_{s}c}{%
Z_{s}e}\left( u^{\prime }\mathbf{b}^{\prime }+\mathbf{V}_{eff}^{\prime
}\right) ,
\end{eqnarray}%
in the same approximation. It is important to stress that the GK theory can
be performed in principle to arbitrary order in $\varepsilon _{M}$ \cite%
{Catto1978,Littlejohn1979,Littlejohn1981,Littlejohn1983,Dubin1983,Hahm1988,Bernstein1985,Balescu,Meiss1990}%
, thus permitting the explicit determination of $m_{s}^{\prime }$ and the
modified EM potentials as well as the relevant guiding-center canonical
momenta.

\subsection{First integrals of motion and guiding-center adiabatic
invariants for AD plasmas}

The exact integrals of motion and the relevant adiabatic invariants
corresponding respectively to Eqs.(\ref{LAGRANGIAN}) and (\ref{LAGR
girocinetica}) can be immediately recovered. By definition, an adiabatic
invariant $P$ of order $n$ with respect to $\varepsilon _{M}$ is conserved
only in an asymptotic sense, i.e., in the sense that $\frac{1}{\Omega
_{cs}^{\prime }}\frac{d}{dt}\ln P=0+O(\varepsilon _{M}^{n+1})$, where $n\geq
0$ is a suitable integer. First we notice that, under the assumptions of
axisymmetry and of Eq.(\ref{b0}), the only first integral of motion is the
canonical momentum $p_{\varphi s}\equiv \frac{\partial \mathcal{L}_{s}}{%
\partial \overset{\cdot }{\varphi }}$ conjugate to the ignorable azimuthal
angle $\varphi $:%
\begin{equation}
p_{\varphi s}=M_{s}R\mathbf{v\cdot e}_{\varphi }+\frac{Z_{s}e}{c}\psi \equiv
\frac{Z_{s}e}{c}\psi _{\ast s}.  \label{p_fi}
\end{equation}%
Since the azimuthal angle $\varphi $ is ignorable also for the GK Lagrangian
$\mathcal{L}_{s}^{\prime }$, it follows that the quantity $p_{\varphi
s}^{\prime }\equiv \frac{\partial \mathcal{L}_{s}^{\prime }}{\partial
\overset{\cdot }{\varphi }}$ is an adiabatic invariant of the prescribed
order, according to the accuracy of the GK transformation used to evaluate $%
\mathcal{L}_{s}^{\prime }$. We shall refer to $p_{\varphi s}^{\prime }$ as
the \textit{guiding-center canonical momentum}. In particular, correct to $%
O(\varepsilon _{M}^{k})$, with $k\geq 1$, one obtains%
\begin{equation}
p_{\varphi s}^{\prime }\equiv \frac{M_{s}}{B^{\prime }}\left( u^{\prime
}I^{\prime }-\frac{c\nabla ^{\prime }\psi ^{\prime }\cdot \nabla ^{\prime
}\Phi _{s}^{^{\prime }eff}}{B^{\prime }}\right) +\frac{Z_{s}e}{c}\psi
^{\prime },  \label{p_fi-HAT}
\end{equation}%
which is an adiabatic invariant of $O(\varepsilon _{M}^{k+1})$, with $k\geq
1 $. Furthermore, the total particle energy
\begin{equation}
\left. E_{s}=\frac{M_{s}}{2}v^{2}\mathbf{+}{Z_{s}e}\Phi _{s}^{eff}(\mathbf{x}%
,\varepsilon _{M}^{n}t),\right.  \label{total_energy}
\end{equation}%
with $n\geq 1$, and the GK Hamiltonian $\mathcal{H}_{s}^{\prime }$ are also
adiabatic invariants of order $n$. Finally, in GK theory, by construction,
the momentum $p_{\phi ^{\prime }s}^{\prime }=\partial \mathcal{L}%
_{s}^{\prime }/\partial \overset{\cdot }{\phi ^{\prime }}$ conjugate to the
gyrophase, as well as the related magnetic moment $m_{s}^{\prime }$ defined
as$\ m_{s}^{\prime }\equiv \frac{Z_{s}e}{M_{s}c}p_{\phi ^{\prime }s}^{\prime
},$\ are adiabatic invariants. As shown by Kruskal (1962 \cite{Kruskal}) it
is always possible to determine $\mathcal{L}_{s}^{\prime }$ so that $%
m_{s}^{\prime }$ is an adiabatic invariant of arbitrary order in $%
\varepsilon _{M}$. In particular, the leading-order approximation is $%
m_{s}^{\prime }\cong \mu _{s}^{\prime }\equiv \frac{M_{s}w^{\prime 2}}{%
2B^{\prime }}$.

Note that the allowance of slow time variations for $E_{s}$ is an elementary
consequence of assumption (\ref{b0}), which allows us to describe realistic
configurations of AD plasmas which slowly evolve in time.

\subsection{Particle trapping phenomena}

GK theory permits explicit treatment of particle trapping corresponding to
the existence of forbidden regions for the motion of charged particles
arising from conservation of energy and magnetic moment. Conservation of the
guiding-center Hamiltonian (\ref{HAM giricinetica}) and the magnetic moment $%
\mu_{s}^{\prime}$ (leading-order approximation) give rise to some
implications. Combining the two identities to express the parallel velocity $%
u^{\prime }$, and using the definition (\ref{V^drift}), we find%
\begin{equation}
u^{\prime }=\pm \sqrt{\frac{2}{M_{s}}\left[ \mathcal{H}_{s}^{\prime (1)}-\mu
_{s}^{\prime }B^{\prime }-{Z_{s}e}\Phi _{s}^{\prime eff}-\frac{M_{s}}{2}%
V_{eff}^{\prime 2}\right] }.  \label{trap}
\end{equation}%
Therefore $u^{\prime }$ is a local function of the guiding-center position
vector $\mathbf{x}^{\prime }$ and, thanks to axisymmetry, of the
corresponding flux coordinates ($\psi ^{\prime },\vartheta ^{\prime }$).
Since the argument of the square root must be non-negative, this means that $%
u^{\prime }$ is only defined in the subset of the configuration space
spanned by ($\psi ^{\prime },\vartheta ^{\prime }$) where this property
holds. It follows that if the argument becomes null for given $\mathcal{H}%
_{s}^{\prime (1)}$ and $\mu _{s}^{\prime }$, the parallel velocity must
change sign so that the particle undergoes a spatial reflection. The points
of the configuration space where this occurs are the so-called \textit{%
mirror points}. The existence of these points may generate various kinetic
phenomena in AD plasmas. In particular, for open magnetic surfaces,
particles can in principle experience zero, one or two reflections
corresponding respectively to \textit{passing particles (PPs), bouncing
particles (BPs) }and \textit{trapped particles (TPs)}. In the present case,
since the right hand side of Eq.(\ref{trap}) depends on the magnitude of the
magnetic field ($B^{\prime }$), the effective potential energy (${Z_{s}e}%
\Phi _{s}^{\prime eff}$) and the centrifugal potential ($\frac{M_{s}}{2}%
V_{eff}^{\prime 2}$), we shall refer to the TPs case as \emph{gravitational
EM trapping}. In Section VII we shall investigate some consequences of
trapping phenomena for the dynamics of ADs.

\section{Construction of the QSA-KDF: generalized solution}

In this section we show that the equilibrium generalized bi-Maxwellian
solution for the KDF obtained in Paper I can be extended to QSA-KDFs
describing axisymmetric AD plasmas with the following features:

1) The KDF is also axisymmetric;

2) Each species in the collisionless plasma is considered to be associated
with a suitable set of \textit{sub-species} (referring to the different
populations mentioned above), each one having a different KDF;

3) Temperature anisotropy: for all of the species, it is assumed that
different parallel and perpendicular temperatures are allowed (with respect
to the local direction of the magnetic field);

4) Accretion flow velocity: a non-vanishing species dependent poloidal flow
velocity is prescribed;

5) Open, locally nested magnetic flux surfaces: the magnetic field is taken
to allow quasi-stationary solutions with magnetic flux lines belonging to
open and locally nested magnetic surfaces;

6) Kinetic constraints: suitable functional dependencies are imposed so that
the KDF is an adiabatic invariant;

7) Analytic form: the solution is required to be asymptotically
\textquotedblleft close\textquotedblright\ to a local bi-Maxwellian in order
to permit comparisons with previous literature dealing with Maxwellian or a
bi-Maxwellian KDFs (see for example \cite{Quataert2002, Quataert2007B,
Snyder1997}).

Requirement 2) is suggested by observations of collisionless plasmas. For
example, in the solar wind plasma both ion and electron species are
described by superpositions of shifted bi-Maxwellian distributions.
Requirements 1) - 7) clearly imply that the solution cannot generally be a
Maxwellian. However, in analogy with Paper I, it is possible to show that
they can be fulfilled by a suitable modified bi-Maxwellian expressed solely
in terms of first integrals of motion and adiabatic invariants \cite%
{Cremasch2008-2,Cremasch2009,Kocha10}. It follows that this is necessarily a
QSA-KDF. A set of fluid equations can then readily be determined using this
solution, expressed in terms of four moments of the KDF [corresponding to
the species number density, flow velocity and the parallel and perpendicular
temperatures]. These equations which, by construction, satisfy a kinetic
closure condition, are also useful for comparing with previous fluid
treatments.

For consistency with the notation of Paper I, we again use the symbol
\textquotedblleft $\wedge $ \textquotedblright\ to denote physical
quantities which refer to the treatment of anisotropic temperatures, unless
otherwise specified, but in the present work, for greater generality, the
symbol \textquotedblleft $\ast $\textquotedblright\ is used to denote
variables which depend on both the canonical momentum $\psi _{\ast s}$ and
the total particle energy $E_{s}$.

In line with all of the previous requirements, it is possible to show that a
particular solution for the QSA-KDF is given by:%
\begin{eqnarray}
\widehat{f_{\ast s}} &=&\frac{\widehat{\beta _{\ast s}}}{\left( 2\pi
/M_{s}\right) ^{3/2}\left( T_{\parallel \ast s}\right) ^{1/2}}  \label{sol1}
\\
&&\times \exp \left\{ -\frac{K_{\ast s}}{T_{\parallel \ast s}}-m_{s}^{\prime
}\widehat{\alpha _{\ast s}}\right\} ,  \notag
\end{eqnarray}%
which we refer to as the \emph{Generalized bi-Maxwellian KDF} \emph{with
parallel velocity perturbations}. Here $\widehat{f_{\ast s}}$ is defined in
the phase-space $\Gamma =\Gamma _{r}\times \Gamma _{u}$, where $\Gamma _{r}$
and $\Gamma _{u}$ are both identified with suitable subsets of the Euclidean
space $%
\mathbb{R}
^{3}$. The notation is as follows:%
\begin{eqnarray}
\widehat{\beta _{\ast s}} &\equiv &\frac{\eta _{s}}{\widehat{T}_{\perp s}},
\\
\widehat{\alpha _{\ast s}} &\equiv &\frac{B^{\prime }}{\widehat{\Delta
_{T_{s}}}}, \\
K_{\ast s} &\equiv &E_{s}-\ell _{\varphi s}\varpi _{\ast s},
\end{eqnarray}%
with $E_{s}$ and $\psi _{\ast s}$ given by Eqs.(\ref{total_energy}) and (\ref%
{p_fi}) respectively, while $\frac{1}{\widehat{\Delta _{T_{s}}}}\equiv \frac{%
1}{\widehat{T}_{\perp s}}-\frac{1}{T_{\parallel \ast s}}$. By construction, $%
\ell _{\varphi s}$ has the dimensions of an angular momentum, while $\varpi
_{\ast s}$ has those of a frequency. In contrast with the solution obtained
in Paper I, $\varpi _{\ast s}$ is not necessarily associated here with a
purely azimuthal leading-order velocity. In general $K_{\ast s}$ can, in
fact, be represented as%
\begin{equation}
K_{\ast s}=E_{s}-\frac{Z_{s}e}{c}\psi _{\ast s}\Omega _{\ast s}-p_{\varphi
s}^{\prime }\xi _{\ast s}=H_{\ast s}-p_{\varphi s}^{\prime }\xi _{\ast s}.
\label{kstar}
\end{equation}%
Here $H_{\ast s}\equiv E_{s}-\frac{Z_{s}e}{c}\psi _{\ast s}\Omega _{\ast s}$
has the same meaning as the analogous quantity used in Paper I, with $\Omega
_{\ast s}$ being related to the azimuthal rotational frequency. In Eq.(\ref%
{kstar}) $\xi _{\ast s}$ is a frequency associated with the leading-order
guiding-center canonical momentum $p_{\varphi s}^{\prime }$ defined in Eq.(%
\ref{p_fi-HAT}), which is an adiabatic invariant depending on $u^{\prime }$
and, by definition, is independent of the gyrophase angle. As we shall show
at the end of this section, this feature can be used to require that the
QSA-KDF carries a non-vanishing parallel flow velocity. This can be related
to a net accretion flow arising in the AD plasma. Finally, by substituting
Eq.(\ref{kstar}) into Eq.(\ref{sol1}) we reach the equivalent representation
for the QSA-KDF:%
\begin{eqnarray}
\widehat{f_{\ast s}} &=&\frac{\widehat{\beta _{\ast s}}}{\left( 2\pi
/M_{s}\right) ^{3/2}\left( T_{\parallel \ast s}\right) ^{1/2}}  \label{sol2}
\\
&&\times \exp \left\{ -\frac{H_{\ast s}}{T_{\parallel \ast s}}+\frac{%
p_{\varphi s}^{\prime }\xi _{\ast s}}{T_{\parallel \ast s}}-m_{s}^{\prime }%
\widehat{\alpha _{\ast s}}\right\} .  \notag
\end{eqnarray}%
In order for the solution (\ref{sol2}) [or equivalently (\ref{sol1})] to be
a function of the integrals of motion and of the adiabatic invariants, the
functions $\left\{ \Lambda _{\ast s}\right\} \equiv \left\{ \widehat{\beta
_{\ast s}},\widehat{\alpha _{\ast s}},T_{\parallel \ast s},\Omega _{\ast
s},\xi _{\ast s}\right\} $, which we will refer to as \emph{structure
functions}, must be adiabatic invariants by themselves. To further
generalize the solution of Paper I, we shall here retain a functional
dependence on both the total particle energy and the canonical momentum,
thus imposing the functional dependencies%
\begin{equation}
\Lambda _{\ast s}=\Lambda _{\ast s}\left( \psi _{\ast s},E_{s}\right) ,
\label{kincon}
\end{equation}%
which will be referred to in the following as \emph{kinetic constraints}.
The kinetic constraints (\ref{kincon}) provide the most general solution for
$\widehat{f_{\ast s}}$. It can be shown that the physical motivation behind
imposing these dependencies lies essentially in the fact that the asymptotic
condition of small inverse aspect ratio (adopted previously in Paper I) is
no longer valid. In the present context, the kinetic solution is no longer
restricted to localized spatial domains in the disk but applies to the
general configuration of open magnetic surfaces. This in turn implies that
the structure functions are generally not simply flux-functions on the
magnetic surfaces. In previous treatments (Paper I and Ref.\cite{Catto1987}%
), the structure functions were identified with $\left\{ \widehat{\beta
_{\ast s}},\widehat{\alpha _{\ast s}},T_{\parallel \ast s},\Omega _{\ast
s}\right\} $ and $\left\{ \widehat{\beta _{\ast s}}\equiv N_{\ast s},T_{\ast
s}\right\} $ for non-isotropic and isotropic generalized Maxwellian KDFs
respectively.

Some basic properties of $\widehat{f_{\ast s}}$ are:

Property 1: $\widehat{f_{\ast s}}$ is itself an adiabatic invariant, and is
therefore an asymptotic solution of the stationary Vlasov equation, i.e., a
QSA-KDF;

Property 2: $\widehat{f_{\ast s}}$ is only defined in the subset of
phase-space where the adiabatic invariants $p_{\varphi s}^{\prime }$, $%
\mathcal{H}_{s}^{\prime (1)}$ and $m_{s}^{\prime }$ are defined. It follows
that $\widehat{f_{\ast s}}$ is suitable for describing both circulating and
trapped particles (see the related discussion in Section 7);

Property 3: all of the velocity-moment equations obtained from the Vlasov
equation (and in particular the continuity and linear momentum fluid
equations) are identically satisfied in an asymptotic sense, i.e.,
neglecting corrections of $O\left( \varepsilon _{M}^{n+1}\right) $;

Property 4: its velocity moments, to be identified with the fluid fields,
are unique once $\widehat{f_{\ast s}}$ is prescribed in terms of the
structure functions;

Property 5: it generalizes the solution earlier presented in Paper I: a) by
using both $p_{\varphi s}^{\prime }$ and $m_{s}^{\prime }$ as adiabatic
invariants and b) because of the new kinetic constraints.

It follows immediately that the solution (\ref{sol2}) does indeed carry
finite parallel velocity perturbations. Invoking the definitions (\ref%
{p_fi-HAT}) and (\ref{kstar}), Eq.(\ref{sol2}) can be re-written as%
\begin{eqnarray}
&&\left. \widehat{f_{\ast s}}=\frac{\widehat{\beta _{\ast s}}\exp \left[
\frac{X_{\ast s}}{T_{\parallel \ast s}}\right] }{\left( 2\pi /M_{s}\right)
^{3/2}\left( T_{\parallel \ast s}\right) ^{1/2}}\right.  \label{sol3} \\
&&\times \exp \left\{ -\frac{M_{s}\left( \mathbf{v}-\mathbf{V}_{\ast
s}-U_{\parallel \ast s}^{\prime }\mathbf{b}^{\prime }\right) ^{2}}{%
2T_{\parallel \ast s}}-m_{s}^{\prime }\widehat{\alpha _{\ast s}}\right\} ,
\notag
\end{eqnarray}%
where $\mathbf{V}_{\ast s}=\mathbf{e}_{\varphi }R\Omega _{\ast s}\left( \psi
_{\ast s},E_{s}\right) $ and%
\begin{equation}
X_{\ast s}\equiv M_{s}\frac{\left\vert \mathbf{V}_{\ast s}\right\vert ^{2}}{2%
}+\frac{Z_{s}e}{c}\psi \Omega _{\ast s}-Z_{s}e\Phi _{s}^{eff}+\Upsilon
_{\ast s}^{\prime }.
\end{equation}%
Here the function $\Upsilon _{\ast s}^{\prime }$ is defined as
\begin{eqnarray}
\Upsilon _{\ast s}^{\prime } &\equiv &\frac{M_{s}U_{\parallel \ast
s}^{\prime 2}}{2}\left( 1+\frac{2\Omega _{\ast s}}{\xi _{\ast s}}\right) + \\
&&-\left( \frac{M_{s}c\nabla ^{\prime }\psi ^{\prime }\cdot \nabla ^{\prime
}\Phi _{s}^{\prime eff}}{B^{\prime 2}}-\frac{Z_{s}e}{c}\psi ^{\prime
}\right) \xi _{\ast s},  \notag
\end{eqnarray}%
with $U_{\parallel \ast s}^{\prime }=\frac{I^{\prime }}{B^{\prime }}\xi
_{\ast s}\left( \psi _{\ast s},E_{s}\right) $. Note that $U_{\parallel \ast
s}^{\prime }$ is non-zero only if the toroidal magnetic field is
non-vanishing. This quantity is independent of $\mathbf{V}_{\ast s}$ and is
clearly associated with a parallel flow velocity (i.e., having both poloidal
and toroidal components), referred to here as a \textit{parallel velocity
perturbation}. This perturbation enters the solution via the adiabatic
invariant $p_{\varphi s}^{\prime }$ and therefore its inclusion is
consistent with the requirement that KDF is an adiabatic invariant.

Finally we note that the same kinetic constraints (\ref{kincon}) also apply
to the solution (\ref{sol3}). However, the functions $\widehat{\beta _{\ast
s}}\exp \left[ \frac{X_{\ast s}}{T_{\parallel \ast s}}\right] ,$ $\mathbf{V}%
_{\ast s},$ $U_{\parallel \ast s}^{\prime }$ and $T_{\parallel \ast s}$
cannot be directly regarded as \emph{fluid fields}, since they still depend
on the single particle velocity via the canonical momentum $\psi _{\ast s}$
and the particle energy $E_{s}$.

\section{Analytical expansion}

Based on Properties 1-5, in this section we determine an approximate
analytical expression for $\widehat{f_{\ast s}}$ obtained by means of
suitable asymptotic expansions. These are carried out in terms of the
following two dimensionless parameters:

1) $\varepsilon _{s}$: which is related to the canonical momentum $\psi
_{\ast s}$. This is defined as (cf Paper I) $\varepsilon _{s}\equiv
\left\vert \frac{L_{\varphi s}}{p_{\varphi s}-L_{\varphi s}}\right\vert
=\left\vert \frac{M_{s}Rv_{\varphi }}{\frac{Z_{s}e}{c}\psi }\right\vert $,
where $v_{\varphi }\equiv \mathbf{v\cdot e}_{\varphi }$ and $L_{\varphi s}$
denotes the species particle angular momentum. We refer to the AD plasma as
being \emph{strongly magnetized} if $0<\varepsilon _{s}\ll 1$;

2) $\sigma _{s}$: which is related to the total particle energy $E_{s}$.
This is defined as $\sigma _{s}\equiv \left\vert \frac{\frac{M_{s}}{2}v^{2}}{%
{Z_{s}e}\Phi _{s}^{eff}}\right\vert $, i.e., it is the ratio between the
kinetic energy and potential energy of the particle. For bound orbits $%
E_{s}<0$, and so $\sigma _{s}<1$.

In the following, we treat $\varepsilon _{s}$ and $\sigma _{s}$ as
infinitesimals of the same order, with $\varepsilon _{s}\sim \sigma _{s}\ll
1 $ and then $\varepsilon _{s}$ and $\sigma _{s}$ can be used for performing
a Taylor expansion of the implicit dependencies contained in the structure
functions by setting $\psi _{\ast s}\cong \psi +O\left( \varepsilon
_{s}\right) $ and $E_{s}\cong {Z_{s}e}\Phi _{s}^{eff}+O\left( \sigma
_{s}\right) $ to leading order. This implies that the linear asymptotic
expansion for the structure functions, obtained neglecting corrections of $%
O\left( \varepsilon _{s}\sigma _{s}\right) ,$ as well as of $O\left(
\varepsilon _{s}^{k}\right) $ and $O\left( \sigma _{s}^{k}\right) $, with $%
k\geq 2$, is%
\begin{eqnarray}
\Lambda _{\ast s} &\cong &\Lambda _{s}+\left( \psi _{\ast s}-\psi \right)
\left[ \frac{\partial \Lambda _{\ast s}}{\partial \psi _{\ast s}}\right]
_{\substack{ \psi _{\ast s}=\psi  \\ E_{s}={Z_{s}e}\Phi _{s}^{eff}}}+  \notag
\\
&&+\left( E_{s}-{Z_{s}e}\Phi _{s}^{eff}\right) \left[ \frac{\partial \Lambda
_{\ast s}}{\partial E_{s}}\right] _{\substack{ \psi _{\ast s}=\psi  \\ E_{s}=%
{Z_{s}e}\Phi _{s}^{eff}}},  \label{espan0}
\end{eqnarray}%
where
\begin{equation}
\Lambda _{s}\equiv \left. \Lambda _{\ast s}\right\vert _{\substack{ \psi
_{\ast s}=\psi  \\ E_{s}={Z_{s}e}\Phi _{s}^{eff}}}.
\end{equation}%
To perform the corresponding expansion for $\widehat{f_{\ast s}}$, we leave
unchanged the dependence in terms of the guiding-center canonical momentum $%
p_{\varphi s}^{\prime }$, while retaining the leading-order approximation
for the magnetic moment only in the linear perturbation terms of Eq.(\ref%
{espan0}). Then, it is straightforward to prove that for strongly magnetized
and bound plasmas, the following relation holds to leading-order:
\begin{equation}
\widehat{f_{\ast s}}\cong \widehat{f_{s}}\left( p_{\varphi s}^{\prime
},m_{s}^{\prime }\right) \left[ 1+h_{Ds}^{1}+h_{Ds}^{2}\right] ,
\label{solo}
\end{equation}%
where $h_{Ds}^{1}$ and $h_{Ds}^{2}$ represent the so-called \emph{%
diamagnetic parts} of $\widehat{f_{\ast s}}$ (see the definition below). The
definitions are then as follows:

\emph{First}, the leading-order distribution $\widehat{f_{s}}\left(
p_{\varphi s}^{\prime },m_{s}^{\prime }\right) $ is expressed as%
\begin{eqnarray}
&&\left. \widehat{f_{s}}\left( p_{\varphi s}^{\prime },m_{s}^{\prime
}\right) =\frac{n_{s}}{\left( 2\pi /M_{s}\right) ^{3/2}\left( T_{\parallel
s}\right) ^{1/2}T_{\perp s}}\right.  \notag \\
&&\times \exp \left\{ -\frac{M_{s}\left( \mathbf{v}-\mathbf{V}%
_{s}-U_{\parallel s}^{\prime }\mathbf{b}^{\prime }\right) ^{2}}{%
2T_{\parallel s}}-m_{s}^{\prime }\frac{B^{\prime }}{\Delta _{T_{s}}}\right\}
\label{solo3}
\end{eqnarray}%
which we will here call the \emph{bi-Maxwellian KDF with parallel velocity
perturbations}. Here $\frac{1}{\Delta _{T_{s}}}\equiv \frac{1}{T_{\perp s}}-%
\frac{1}{T_{\parallel s}}$ is related to the temperature anisotropy, the
number density is defined as%
\begin{equation}
n_{s}=\eta _{s}\exp \left[ \frac{X_{s}}{T_{\parallel s}}\right]
\end{equation}%
and%
\begin{equation}
X_{s}\equiv \left( M_{s}\frac{R^{2}\Omega _{s}^{2}}{2}+\frac{Z_{s}e}{c}\psi
\Omega _{s}-Z_{s}e\Phi _{s}^{eff}+\Upsilon _{s}^{\prime }\right) ,
\end{equation}%
with $\eta _{s}$ denoting the \emph{pseudo-density}. The function $\Upsilon
_{s}^{\prime }$ is defined as
\begin{eqnarray}
\Upsilon _{s}^{\prime } &\equiv &\frac{M_{s}U_{\parallel s}^{\prime 2}}{2}%
\left( 1+\frac{2\Omega _{s}}{\xi _{s}^{\prime }}\right) +  \notag \\
&&-\left( \frac{M_{s}c\nabla ^{\prime }\psi ^{\prime }\cdot \nabla ^{\prime
}\Phi _{s}^{\prime eff}}{B^{\prime 2}}-\frac{Z_{s}e}{c}\psi ^{\prime
}\right) \xi _{s}.
\end{eqnarray}%
Note that $\mathbf{V}_{s}=\mathbf{e}_{\varphi }R\Omega _{s}$ and $%
U_{\parallel s}^{\prime }=\frac{I^{\prime }}{B^{\prime }}\xi _{s}$ define,
respectively, the leading-order azimuthal flow velocity and the
leading-order parallel velocity perturbation of the fluid. Then, the
following kinetic\emph{\ }constraints are implied from (\ref{kincon}), to
leading-order, for the structure functions:%
\begin{equation}
\Lambda _{s}=\Lambda _{s}\left( \psi ,Z_{s}e\Phi _{s}^{eff}\right) .
\label{kinkin}
\end{equation}

\emph{Second}, the diamagnetic parts $h_{Ds}^{1}$ and $h_{Ds}^{2}$ of $%
\widehat{f_{\ast s}}$, due respectively to the expansions of the canonical
momentum and the total energy, are given by%
\begin{eqnarray}
h_{Ds}^{1} &=&\left\{ \frac{cM_{s}R}{Z_{s}e}\left[ Y_{1}+Y_{3}\right] +\frac{%
M_{s}R}{T_{\parallel s}}Y_{2}\right\} \left( \mathbf{v\cdot }\widehat{e}%
_{\varphi }\right) ,  \label{hd1} \\
h_{Ds}^{2} &=&\frac{M_{s}}{2Z_{s}e}\left\{ Y_{4}-\frac{Z_{s}e}{T_{\parallel
s}}Y_{5}+\frac{p_{\varphi s}^{\prime }\xi _{s}}{T_{\parallel s}}%
C_{5s}\right\} v^{2}.  \label{hd2}
\end{eqnarray}%
Here $Y_{i},$ $i=1,5$, is defined as
\begin{eqnarray}
Y_{1} &\equiv &\left[ A_{1s}+A_{2s}\left( \frac{H_{s}}{T_{\parallel s}}-%
\frac{1}{2}\right) -\mu _{s}^{\prime }\widehat{A_{4s}}\right] , \\
Y_{2} &\equiv &\Omega _{s}\left[ 1+\psi A_{3s}\right] , \\
Y_{3} &\equiv &\left[ \frac{p_{\varphi s}^{\prime }\xi _{s}}{T_{\parallel s}}%
A_{5s}-A_{2s}\frac{p_{\varphi s}^{\prime }\xi _{s}}{T_{\parallel s}}\right] ,
\\
Y_{4} &\equiv &\left[ C_{1s}+C_{2s}\left( \frac{H_{s}}{T_{\parallel s}}-%
\frac{1}{2}\right) -\mu _{s}^{\prime }\widehat{C_{4s}}\right] , \\
Y_{5} &\equiv &\left[ 1+\frac{\Omega _{s}\psi }{c}C_{3s}\right] ,
\end{eqnarray}%
where $H_{s}={E}_{s}-\frac{Z_{s}e}{c}\psi _{s}\Omega _{s}$ and the following
definitions have been introduced: $A_{1s}\equiv \frac{\partial \ln \beta _{s}%
}{\partial \psi },$ $A_{2s}\equiv \frac{\partial \ln T_{\parallel s}}{%
\partial \psi },$ $A_{3s}\equiv \frac{\partial \ln \Omega _{s}}{\partial
\psi },$ $\widehat{A_{4s}}\equiv \frac{\partial \widehat{\alpha _{s}}}{%
\partial \psi },$ $A_{5s}\equiv \frac{\partial \ln \xi _{s}}{\partial \psi }$
and $C_{1s}\equiv \frac{\partial \ln \beta _{s}}{\partial \Phi _{s}^{eff}},$
$C_{2s}\equiv \frac{\partial \ln T_{\parallel s}}{\partial \Phi _{s}^{eff}},$
$C_{3s}\equiv \frac{\partial \ln \Omega _{s}}{\partial \Phi _{s}^{eff}}$, $%
\widehat{C_{4s}}\equiv \frac{\partial \widehat{\alpha _{s}}}{\partial \Phi
_{s}^{eff}},$ $C_{5s}\equiv \frac{\partial \ln \xi _{s}}{\partial \Phi
_{s}^{eff}}$.

We should make a number of comments here:

1) The functional forms of the leading-order number density, the parallel
and azimuthal flow velocities and the temperatures carried by the
bi-Maxwellian KDF, are naturally determined in terms of $\psi $ and $%
Z_{s}e\Phi_{s}^{eff}$. The effective potential $\Phi _{s}^{eff}$ is
generally a function of the form $\Phi_{s}^{eff}=\Phi _{s}^{eff}(\mathbf{x}%
,\varepsilon_{M}^{k}t),$ with $\mathbf{x}=\left( R,z\right) $, since
generally neither the gravitational potential nor the electrostatic
potential are expected to be flux functions in the present case. Hence, in
magnetic coordinates, it follows that the structure functions are of the
form $\Lambda _{s}\equiv \overline{\Lambda_{s}}\left(\psi ,\vartheta
,\varepsilon _{M}^{k}t\right)$;

2) The coefficients $A_{is}$ and $C_{is}$, $i=1,5$, can be identified with
effective \textit{thermodynamic forces}: $A_{5s}$ carries the contribution
of the parallel velocity perturbation, while the $C_{is}$, $i=1,5$, are due
to the energy dependence contained in the structure functions;

3) We stress that the energy dependence contained in the kinetic constraints
is non trivial and cannot be included simply by redefining the structure
functions (e.g., by transforming the magnetic coordinates). In fact, besides
modifying the leading order structure functions (see point 1 above), it
gives rise to the new diamagnetic contribution $h_{Ds}^{2}$. Eq.(\ref{solo})
is therefore a generalization of the analogous solution obtained in Paper I,
which also appears in standard tokamak transport theory \cite{Catto1987},
where the relevant structure functions were considered solely as flux
functions. Including the effect of the parallel velocity perturbations gives
rise to contributions to $h_{Ds}^{2}$ which are even with respect to $%
u^{\prime }$;

4) In the analytical expansion, we have assumed that the scale-length $L$ is
of the same order in $\varepsilon _{s}$ as the characteristic scale-lengths
associated with the structure functions;

5) We have performed the analysis distinguishing between the different
plasma species. Since this is an asymptotic estimation, the analytical
expansion can be different for ions and electrons, particularly for the
terms appearing in the diamagnetic part, depending on the relative
magnitudes of the parameters $\varepsilon _{s}$ and $\sigma _{s}$. On the
other hand, because of the double expansion and the energy dependence, the
asymptotic solution for the two species can hold also in different spatial
domains;

6) The KDF $\widehat{f_{s}}\left( p_{\varphi
s}^{\prime},m_{s}^{\prime}\right)$ also satisfies Property 2: namely, it is
only defined in the subset of phase-space where the parallel velocity $%
\left\vert u^{\prime}\right\vert $ is a real function. It is therefore
suitable for properly describing particle trapping;

7) Finally, we stress that the QSA-KDF (\ref{sol2}) obtained here, reduces
asymptotically to the expression reported in previous paper (see Eq.(10) in
Paper I) when the following conditions are satisfied: a) parallel velocity
perturbations are ignored, namely the structure function $\xi _{\ast s}$ is
set to zero; b) closed nested magnetic surfaces are considered; c) large
aspect ratio ordering, $1/\delta \gg 1$, is invoked (see the definition in
Paper I). In this case, the effective potential is solely a flux-function to
leading order, while the diamagnetic contribution $h_{Ds}^{2}$ can be shown
to be of higher order than $h_{Ds}^{1}$.

\bigskip

\section{Moment equations}

In this section we discuss the connection between the kinetic
treatment presented here and the corresponding fluid approach,
obtained by describing the plasma in terms of a suitable set of
fluid fields. The latter can in principle be specified as required
by experimental observations and identified with the relevant
physical observables. Important practical aspects of the present
theory concern the explicit evaluation of the fluid fields
associated with the QSA-KDF, and the conditions for validity of
the relevant moment equations.

For definiteness, let us require that:

\begin{enumerate}
\item The KDF, the EM fields $\left\{
\mathbf{E},\mathbf{B}\right\} $ and the corresponding EM
potentials $\left\{ \Phi ,\mathbf{A}\right\} $ are all exactly
axisymmetric and, moreover, stationary in an asymptotic sense,
i.e. neglecting corrections of $O(\varepsilon _{M}^{n+1})$;

\item The KDF is identified with\textit{\ }the QSA-KDF $\widehat{f_{\ast s}}%
\left( E_{s},\psi _{\ast s},m_{s}^{\prime }\right) $ which, by
assumption, is required to be an adiabatic invariant of
$O(\varepsilon _{M}^{n+1})$. By construction $\widehat{f_{\ast
s}}\left( E_{s},\psi _{\ast s},m_{s}^{\prime
}\right) $ is a solution of the \textit{asymptotic Vlasov equation}%
\begin{equation}
\frac{1}{\Omega _{cs}^{\prime }}\frac{d}{dt}\ln \widehat{f_{\ast s}}%
=0+O\left( \varepsilon _{M}^{n+1}\right) .  \label{asvla}
\end{equation}%
This equation holds by definition up to infinitesimals of $O\left(
\varepsilon _{M}^{n+1}\right) $, where $n$ is an arbitrary positive integer;

\item The magnetic field is taken to be of the form (\ref{B FIELD}).
\end{enumerate}

As a basic consequence of these assumptions, the stationary fluid equations
following from the Vlasov equation are necessarily all identically satisfied
in an asymptotic sense, i.e., again neglecting corrections of $O\left(
\varepsilon _{M}^{n+1}\right) $. In fact if $Z(\mathbf{x})$ is an arbitrary
weight function, identified for example with $Z=\left( 1,\mathbf{v}%
,v^{2}\right) ,$ then the generic moment of Eq.(\ref{asvla}) is:%
\begin{equation}
\int_{\Gamma _{u}}d^{3}vZ\frac{d}{dt}\widehat{f_{\ast s}}=0+O\left(
\varepsilon _{M}^{n+1}\right) ,  \label{dd}
\end{equation}%
where $\Gamma _{u}$ denotes the appropriate velocity space of integration.
Using the chain rule, this can be written as%
\begin{eqnarray}
&&\left. \int_{\Gamma _{u}}d^{3}vZ\left\{ \frac{d\psi _{\ast }}{dt}\frac{%
\partial \widehat{f_{\ast s}}}{\partial \psi _{\ast }}+\frac{dE_{s}}{dt}%
\frac{\partial \widehat{f_{\ast s}}}{\partial E_{s}}+\frac{dm_{s}^{\prime }}{%
dt}\frac{\partial \widehat{f_{\ast s}}}{\partial m_{s}^{\prime }}\right\}
=\right.  \notag \\
&&\left. =0+O\left( \varepsilon _{M}^{n+1}\right) .\right.
\end{eqnarray}%
On the other hand, Eq.(\ref{dd}) can also be represented as%
\begin{equation}
\int_{\Gamma _{u}}d^{3}v\left\{ \frac{d}{dt}\left[ Z\widehat{f_{\ast s}}%
\right] -\widehat{f_{\ast s}}\frac{d}{dt}Z\right\} =0+O\left( \varepsilon
_{M}^{n+1}\right) ,
\end{equation}%
which recovers the usual form of the velocity-moment equations in terms of
suitable (and \textit{uniquely defined}) fluid fields. For $Z=\left( 1,%
\mathbf{v}\right) $ one obtains, in particular, that the species
continuity and linear momentum fluid equations are satisfied
identically up to infinitesimals of $O\left( \varepsilon
_{M}^{n+1}\right) :$

\begin{equation}
\nabla \cdot \left( n_{s}^{tot}\mathbf{V}_{s}^{tot}\right) =0+O\left(
\varepsilon _{M}^{n+1}\right) ,  \label{moment1}
\end{equation}%
\begin{eqnarray}
&&\left. M_{s}\mathbf{V}_{s}^{tot}\cdot \nabla \mathbf{V}_{s}^{tot}+\nabla
\cdot \underline{\underline{\Pi }}_{s}^{tot}+Z_{s}en_{s}^{tot}\nabla \Phi
_{s}^{eff}+\right. \ \ \ \ \ \ \   \notag \\
&&\left. -\frac{Z_{s}e}{c}\mathbf{V}_{s}^{tot}\times \mathbf{B}=0+O\left(
\varepsilon _{M}^{n+1}\right) .\right.  \label{moment2}
\end{eqnarray}%
Similarly, the law of conservation of the species total canonical momentum
can be recovered by setting $Z=\psi _{\ast s}$, namely%
\begin{equation}
\int_{\Gamma _{u}}d^{3}v\frac{d}{dt}\left[ \psi _{\ast s}\widehat{f_{\ast s}}%
\right] =0+O\left( \varepsilon _{M}^{n+1}\right) .
\end{equation}%
In the stationary case this implies the \textit{species angular momentum
conservation law}%
\begin{equation}
\nabla \cdot \left[ R^{2}\underline{\underline{\Pi }}_{s}^{tot}\cdot \nabla
\varphi +\mathbf{V}_{s}^{tot}L_{s}^{tot}\right] +\frac{Z_{s}e}{c}\nabla \psi
\cdot n_{s}^{tot}\mathbf{V}_{s}^{tot}=0  \label{angcons}
\end{equation}%
for the species angular momentum%
\begin{equation}
L_{s}^{tot}\equiv M_{s}R^{2}n_{s}^{tot}\mathbf{V}_{s}^{tot}\cdot \nabla
\varphi .
\end{equation}%
Here the notation is standard. In particular the following velocity moments
of the QSA-KDF can be introduced:

a) \textit{species number density}%
\begin{equation}
n_{s}^{tot}\equiv \int_{\Gamma _{u}}d^{3}v\widehat{f_{\ast s}};  \label{fff1}
\end{equation}

b) \textit{species flow velocity}%
\begin{equation}
\mathbf{V}_{s}^{tot}\equiv \frac{1}{n_{s}^{tot}}\int_{\Gamma _{u}}d^{3}v%
\mathbf{v}\widehat{f_{\ast s}};  \label{fff}
\end{equation}

c) \textit{species tensor pressure}%
\begin{equation}
\underline{\underline{\Pi }}_{s}^{tot}\equiv \int_{\Gamma
_{u}}d^{3}vM_{s}\left( \mathbf{v}-\mathbf{V}_{s}^{tot}\right) \left( \mathbf{%
v}-\mathbf{V}_{s}^{tot}\right) \widehat{f_{\ast s}};  \label{fff2}
\end{equation}

d) \textit{species canonical toroidal momentum}%
\begin{equation}
L_{cs}^{tot}\equiv \int_{\Gamma _{u}}d^{3}v\frac{Z_{s}e}{c}\psi _{\ast s}%
\widehat{f_{\ast s}}.
\end{equation}

It is worth remarking here that \textit{the velocity moments are unique once
the QSA-KDF} $\widehat{f_{\ast s}}$ [see Eq.(\ref{sol1})] \textit{is
prescribed in terms of the structure functions} $\Lambda _{\ast s}.$ On the
other hand, as a result of Eqs.(\ref{asvla}) and (\ref{dd}), it follows that
the stationary fluid moments calculated in terms of the QSA-KDF $\widehat{%
f_{\ast s}}$ are identically solutions of the corresponding stationary fluid
moment equations. In particular, imposing the \textit{quasi-neutrality
condition} in the sense%
\begin{equation}
\sum\limits_{s=i,e}Z_{s}en_{s}^{tot}=0+O\left( \varepsilon _{M}^{k}\right)
\label{qneu}
\end{equation}%
with $k\geq 2$, the total fluid canonical toroidal momentum and
the fluid angular momentum necessarily coincide, namely%
\begin{equation}
L^{tot}\equiv \sum\limits_{s=i,e}L_{s}^{tot}\equiv
\sum\limits_{s=i,e}L_{cs}^{tot}.  \label{Ltot}
\end{equation}%
Let us now illustrate explicitly how it is possible to carry out
such a calculation within the present theory. The evaluation of
the previous fluid fields can be made by using the asymptotic
analytical solution of the QSA-KDF $\widehat{f_{\ast s}}$ derived
in the previous section and given by Eq.(\ref{solo}). For example,
adopting this expansion in the limit of strongly magnetized
plasmas, from Eq.(\ref{fff1}) the species number density
becomes%
\begin{equation}
n_{s}^{tot}\cong \int_{\Gamma _{u}}d^{3}v\left\{ \widehat{f_{s}}\left[
1+h_{Ds}^{1}+h_{Ds}^{2}\right] \right\} ,
\end{equation}%
in which the diamagnetic corrections to the bi-Maxwellian KDF $\widehat{f_{s}%
}$ are polynomial functions of the particle velocity. Analogous
expressions can also be obtained in a straightforward way for the
remaining fluid moments. As pointed out in Paper I and
subsequently in Ref.\cite{Catania2}, the expansion procedure for
$\widehat{f_{\ast s}}$ can in principle be performed to higher
order, allowing for the analytical computation of the
corresponding quasi-stationary fluid fields and the determination
of the relevant kinetic closure conditions for the stationary
moment equations. In the present context we stress that the theory
allows the treatment of multiple-species plasmas including, in
particular, particle trapping phenomena. This is taken into
account by proper definition of the velocity sub-space $\Gamma
_{u}$ in which the integrations are performed. In fact, charged
particles in both open and closed configurations can have mirror
points (TPs and BPs) or be PPs, which are free to stream through
the boundaries of the domain. These populations give different
contributions to the relevant fluid fields and therefore require
separate statistical treatments. The explicit calculation of fluid
fields requires also a preliminary \textit{inverse transformation}
representing all quantities in
terms of the actual particle positions (the FLR expansion, see Eq.(\ref%
{trasformazionui guirocinetiche})). This introduces further correction terms
of order $\varepsilon _{M}^{k}$, $k\geq 1$, into the final analytical
expressions. In contrast with the conclusion reached in Paper I, here we
expect these FLR corrections to be non-negligible due to the requirement $%
\varepsilon _{M,s}\sim \varepsilon _{s}$ holding for open-field
configurations.

\bigskip

\section{Slow time-evolution of the axisymmetric QSA-KDF}

In this section we investigate the temporal evolution of the axisymmetric
QSA-KDF, consistent with the assumptions of Section 2 and the results of
Sections 3 and 4. Two different issues must be addressed: giving an estimate
of the maximum time interval over which the QSA-KDF can be regarded as an
asymptotic stationary solution; and determining the solution of the Vlasov
equation for time intervals longer than the equilibrium one.

For our explicit determination of the time evolution of the QSA-KDF, we make
the following assumptions:

1) That the plasma can be treated as a continuous medium in the kinetic
description. This requires that the species kinetic equation holds on time
and spatial scales which are much longer than the corresponding Langmuir
characteristic times and Debye lengths;

2) That we are considering timescales much shorter than the species
characteristic collisional time $\tau _{C}$, so that it is appropriate to
use the Vlasov equation;

3) That the species KDF and the EM fields vary slowly in time and space with
respect to the corresponding Larmor times and radii, so that the GK
description is valid;

4) That the EM and gravitational fields vary slowly in time, in accordance
with Eq.(\ref{b0}), so that the total energy $E_{s}$ is an adiabatic
invariant. In particular, we require:%
\begin{equation}
\frac{d}{dt}E_{s}=Z_{s}e\frac{\partial }{\partial t}\Phi _{s}^{eff}-\frac{%
Z_{s}e}{c}\mathbf{v}\cdot \frac{\partial }{\partial t}\mathbf{A},
\label{dten}
\end{equation}%
which implies that $\tau _{Ls}\frac{d}{dt}\ln E_{s}\sim O\left( \varepsilon
_{M,s}^{n+1}\right) $, with $n\geq 0$. Consistently with the properties of
solution (\ref{sol2}), we take $n=0$ as a specific case. Note that from here
on, $\tau _{Ls}\equiv \frac{1}{\Omega _{cs}^{\prime }}$ will denote the
species characteristic time associated with the Larmor rotation (the \textit{%
Larmor rotation time}). Since $\frac{d}{dt}\Omega _{\ast s} = \frac{dE_{s}}{%
dt}\frac{\partial }{\partial E_{s}}\Omega _{\ast s}$, it follows that%
\begin{equation}
\frac{d}{dt}H_{\ast s}=\frac{d}{dt}E_{s}\left[ 1-\frac{Z_{s}e}{c}\psi _{\ast
s}\frac{\partial }{\partial E_{s}}\Omega _{\ast s}\right] ;
\end{equation}

5) That the magnetic moment $m_{s}^{\prime }$ and the guiding-center
canonical momentum $p_{\varphi s}^{\prime }$ can be taken as adiabatic
invariants of $O\left(\varepsilon _{M,s}^{j}\right) ,$ with $j\geq n$. The
ordering $\tau _{Ls}\frac{d}{dt}\ln p_{\varphi s}^{\prime }\sim O\left(
\varepsilon _{M,s}^{2}\right) $ holds for the leading-order expression for $%
p_{\varphi s}^{\prime }$ adopted here as follows from Eq.(\ref{p_fi-HAT})
and the fact that, by definition, higher-order correction terms, $\Delta
p_{\varphi s}^{\prime }$, to $p_{\varphi s}^{\prime }$ are independent of
the gyrophase angle $\phi^{\prime }$. In fact, denoting by $\mathcal{L}%
_{s}^{\prime (2)}$ the second-order GK Lagrangian, $\Delta p_{\varphi
s}^{\prime }$ can be estimated as $\Delta p_{\varphi s}^{\prime }=\frac{%
\partial }{\partial \overset{\cdot }{\varphi }}\left[ \mathcal{L}%
_{s}^{\prime (2)}-\mathcal{L}_{s}^{\prime (1)}\right] $ where, by
construction, $\mathcal{L}_{s}^{\prime (1)}$ and $\mathcal{L}_{s}^{\prime
(2)}$ are both gyrophase independent. Note that the assumption made here
requires the construction of a higher-order GK theory in order to correctly
determine $m_{s}^{\prime }$ to the required order in the Larmor-radius
expansion.

The time evolution of the QSA-KDF is in principle determined by two
different mechanisms: the explicit time variation of the EM and
gravitational fields, and the time variation of the guiding-center adiabatic
invariants. However, the choice of the orderings in 4) and 5) above, allows
the time dependence produced only by the EM and gravitational fields to be
singled out.

When assumptions 1) - 5) above hold, it follows that $\tau _{Ls}\frac{d}{dt}%
\ln \widehat{f_{\ast s}}=0+O\left( \varepsilon _{M,s}^{n+1}\right) $, with $%
n\geq 0$ being determined by Eq.(\ref{dten}). Then, ignoring higher-order
corrections%
\begin{equation}
\frac{d}{dt}\ln \widehat{f_{\ast s}}=\frac{dE_{s}}{dt}S_{s},
\end{equation}%
where%
\begin{eqnarray}
S_{s} &\equiv &\frac{\partial \ln \widehat{\beta _{\ast s}}}{\partial E_{s}}%
-m_{s}^{\prime }\frac{\partial \widehat{\alpha _{\ast s}}}{\partial E_{s}}+%
\frac{p_{\varphi s}^{\prime }}{T_{\parallel \ast s}}\frac{\partial \xi
_{\ast s}}{\partial E_{s}}+  \notag \\
&&+\left( \frac{H_{\ast s}}{T_{\parallel \ast s}}-\frac{1}{2}+\frac{%
p_{\varphi s}^{\prime }\xi _{\ast s}}{T_{\parallel \ast s}}\right) \frac{%
\partial \ln T_{\parallel \ast s}}{\partial E_{s}}+  \notag \\
&&-\frac{1}{T_{\parallel \ast s}}\left( 1-\frac{Z_{s}e}{c}\psi _{\ast s}%
\frac{\partial \Omega _{\ast s}}{\partial E_{s}}\right) ,  \label{s}
\end{eqnarray}%
and so the solution $\widehat{f_{\ast s}}$ can be regarded as an exact
kinetic equilibrium for all times $t\geq 0$ such that%
\begin{equation}
\tau _{Ls}\ll t\ll t_{\sup }\ll \tau _{C},  \label{timeint1}
\end{equation}%
where $t_{\sup }\equiv \frac{\tau _{Ls}}{\varepsilon _{M,s}^{n+1}}$. Within
the scope of the above assumptions, we now determine the dynamical evolution
equation which describes the slow time-evolution of the QSA-KDF $\widehat{%
f_{\ast s}}$, for time intervals such that $t$ is within
\begin{equation}
t_{\sup }\ll t\ll \tau _{C}.  \label{timeint2}
\end{equation}%
In analogy with Ref.\cite{Catto1987}, we denote by%
\begin{equation}
f_{s}\equiv \widehat{f_{\ast s}}+g_{s}^{\prime }  \label{pos1}
\end{equation}%
the exact solution of the collisionless Vlasov equation, for which $\frac{d}{%
dt}f_{s}=0$. Here $g_{s}^{\prime }$ is referred to as the \textit{reduced KDF%
}. Following the discussion in Ref.\cite{Catto1987}, regarding the
evaluation of $\frac{d}{dt}g_{s}^{\prime }$: it is straightforward to prove
that $g_{s}^{\prime }$ is gyrophase independent, to lowest order, in the
sense that $\frac{\partial g_{s}^{\prime }}{\partial \phi ^{\prime }}=0$.
Therefore, identifying the GK variables with the set $\mathbf{z\equiv }%
\left( \vartheta ^{\prime },\varphi ,p_{\varphi s}^{\prime },\mathcal{H}%
_{s}^{\prime (1)},m_{s}^{\prime },\phi ^{\prime }\right) $, we shall assume
that $g_{s}^{\prime }$ is axisymmetric and of the form $g_{s}^{\prime
}=g_{s}^{\prime }\left( \vartheta ^{\prime },p_{\varphi s}^{\prime },%
\mathcal{H}_{s}^{\prime (1)},m_{s}^{\prime },t\right) $. The gyro-averaged
dynamical equation for $g_{s}^{\prime }$ can then be obtained to next order
by introducing the gyro-average operator $\left\langle ...\right\rangle
_{\phi ^{\prime }}$ defined as%
\begin{equation}
\left\langle ...\right\rangle _{\phi ^{\prime }}\equiv \frac{1}{2\pi }%
\int_{0}^{2\pi }\left( ...\right) d\phi ^{\prime },
\end{equation}%
with the operation being performed while all of the other GK variables are
held fixed \cite{Catto1987}. It follows that, to leading-order, $\frac{d}{dt}%
g_{s}^{\prime }\cong \frac{\partial }{\partial t}g_{s}^{\prime }+\overset{%
\cdot }{\vartheta }^{\prime }\frac{\partial }{\partial \vartheta ^{\prime }}%
g_{s}^{\prime }$, where the time variation of the guiding-center magnetic
coordinate $\vartheta ^{\prime }$ is given by $\overset{\cdot }{\vartheta }%
^{\prime }\cong \overset{\cdot }{\mathbf{r}}^{\prime }\cdot \nabla ^{\prime
}\vartheta ^{\prime }\cong \left[ u^{\prime }\mathbf{b}^{\prime }+\mathbf{V}%
_{eff}^{\prime }\right] \cdot \nabla ^{\prime }\vartheta ^{\prime }$ to
leading-order, with the equation of motion for $\overset{\cdot }{\mathbf{r}}%
^{\prime }$ following from the gyrokinetic Lagrangian [e.g. from the
leading-order Eq.(\ref{LAGR girocinetica})]. Then, consistently with these
assumptions and ignoring higher-order corrections, it is found that the GK
reduced KDF $g_{s}^{\prime }$ obeys the \textit{reduced GK-Vlasov equation}%
\begin{equation}
\frac{\partial }{\partial t}g_{s}^{\prime }+\overset{\cdot }{\vartheta }%
^{\prime }\frac{\partial }{\partial \vartheta ^{\prime }}g_{s}^{\prime
}=-\left\langle \widehat{f_{\ast s}}S_{s}\frac{dE_{s}}{dt}\right\rangle
_{\phi ^{\prime }}.  \label{timeevol2}
\end{equation}%
It follows that, to leading-order%
\begin{equation}
\left\langle \widehat{f_{\ast s}}S_{s}\frac{dE_{s}}{dt}\right\rangle _{\phi
^{\prime }}\cong f_{s}^{\prime }\left\langle S_{s}\frac{dE_{s}}{dt}%
\right\rangle _{\phi ^{\prime }}.  \label{dopo}
\end{equation}%
Denoting $\widehat{f_{\ast s}}\equiv F_{s}\left( \psi _{\ast s},H_{\ast
s},p_{\varphi s}^{\prime },m_{s}^{\prime }\right) $, $f_{s}^{\prime }$ is
then defined as $f_{s}^{\prime }\equiv F_{s}\left( \frac{c}{Z_{s}e}%
p_{\varphi s}^{\prime },\mathcal{H}_{s}^{\prime (1)},p_{\varphi s}^{\prime
},m_{s}^{\prime }\right) $. The remaining gyrophase average in the last
equation can be performed in a straightforward way using Eqs.(\ref{dten})
and (\ref{s}).

Eq.(\ref{timeevol2}) clearly also holds in the time interval (\ref{timeint1}%
), and so it determines the slow time-evolution for all times $\tau_{Ls}\ll
t\ll \tau _{C}$. For consistency, the non-stationary Maxwell equations must
also be solved with the same accuracy. Eq.(\ref{timeevol2}) must be
supplemented by appropriate boundary conditions: for open magnetic surfaces
with boundaries prescribed on a given magnetic surface $\psi =const.$, at $%
\vartheta =\vartheta _{1}$ and $\vartheta =\vartheta _{2}$, with $\vartheta
_{1}<\vartheta _{2}$ and $\vartheta _{1}$, $\vartheta _{2}$ representing the
internal and external boundaries, these are defined respectively either by
prescribing $f_{s}\left( \vartheta _{1}\right) =f_{s}^{(1)}$ or $f_{s}\left(
\vartheta _{2}\right) =f_{s}^{(2)}$ (see Fig.2 for a schematic view of the
configuration geometry and the meaning of the notation). Both $f_{s}^{(1)}$
and $f_{s}^{(2)}$ are necessarily of the form (\ref{pos1}) but their moments
remain arbitrary in principle. As indicated below, this is essential for
making comparisons with experimental observations.

\begin{figure}[tbp]
\centering
\includegraphics[width=3.3in,height=2.3in]{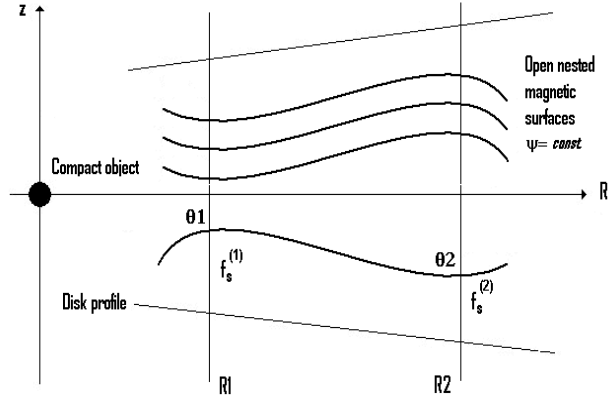} 
\caption{Schematic view of the configuration geometry (not to scale) and
meaning of the notation.}
\end{figure}

The results obtained here have important consequences for the kinetic
description of slow time-evolution of collisionless AD plasmas. Fluid fields
and moment equations can be explicitly determined in terms of Eq.(\ref%
{timeevol2}) by invoking the perturbative expansion outlined in Section 5
and the relations given in Section 6.

\bigskip

\section{The Ampere equation and the kinetic dynamo}

In this section we apply the kinetic solution for the QSA-KDF to
discuss the properties of the Ampere equation and the implications
for the self-generation of magnetic field by the quasi-stationary
AD collisionless plasma. We refer here to this phenomenon as a
\textit{quasi-stationary kinetic dynamo effect}. Generalizing the
treatment presented in Paper I, the Ampere
equation for the self magnetic field becomes:%
\begin{equation}
\nabla \times \mathbf{B}^{self}=\frac{4\pi }{c}\left( \mathbf{J}^{T}+\mathbf{%
J}^{B}+\mathbf{J}^{P}\right) ,  \label{Ampere-2}
\end{equation}%
where $\mathbf{B}^{self}$ has been defined in Eq.(\ref{bself}) and
here we have distinguished between the contributions arising from
PPs, BPs and TPs,
denoting the corresponding total current densities as $\mathbf{J}^{T},%
\mathbf{J}^{B}$ and $\mathbf{J}^{P}$. As described in Sections 5 and 6,
these\ fluid fields can be calculated in closed analytic form to the
required order, by using the asymptotic expansion of the QSA-KDF. This
gives:%
\begin{eqnarray}
\mathbf{J}^{l} &\equiv &\sum\limits_{s=i,e}\mathbf{J}_{s}^{l}=\sum%
\limits_{s=i,e}Z_{s}e\int_{\Gamma _{u}^{l}}d^{3}v\mathbf{v}\widehat{f_{\ast
s}}\cong  \notag \\
&\cong &\sum\limits_{s=i,e}Z_{s}e\int_{\Gamma _{u}^{l}}d^{3}v\mathbf{v}%
\left\{ \widehat{f_{s}}\left[ 1+h_{Ds}^{1}+h_{Ds}^{2}\right] \right\}
\end{eqnarray}%
for $l=T,B,P$ and where $\Gamma _{u}^{l}$ denotes the appropriate velocity
space domain of integration for trapped, bouncing and passing particles
respectively. For convenience of notation, in the following we shall denote as $%
\mathbf{J\equiv J}^{T}+\mathbf{J}^{B}+\mathbf{J}^{P}$ the total
current density entering Eq.(\ref{Ampere-2}). It is possible to
prove that in the case of open magnetic surfaces the total current
density $\mathbf{J}$ in general has non-vanishing components along
all of the three directions identified by the set of magnetic
coordinates $(\psi ,\varphi ,\vartheta )$.
Hence, $\mathbf{J}$ can be represented as%
\begin{equation}
\mathbf{J}=\left( J_{\psi }\nabla \vartheta \times \nabla \varphi
,J_{\varphi }\nabla \varphi ,J_{\vartheta }\nabla \psi \times \nabla \varphi
\right) .
\end{equation}%
Let us now proceed with the study of the Ampere equation. The toroidal
component of Eq.(\ref{Ampere-2}) gives, as usual, the generalized
Grad-Shafranov equation for the poloidal flux function $\psi _{p}$:%
\begin{equation}
\Delta ^{\ast }\psi _{p}=-\frac{4\pi }{c}J_{\varphi },  \label{GSapp}
\end{equation}%
where the elliptic operator $\Delta ^{\ast }$ is defined as $\Delta ^{\ast
}\equiv R^{2}\nabla \cdot \left( R^{-2}\nabla \right) $. The remaining terms
of Eq.(\ref{Ampere-2}) along the directions $\nabla \vartheta \times \nabla
\varphi $ and $\nabla \psi \times \nabla \varphi $ give two equations for
the toroidal component of the magnetic field $I/R$. These are respectively%
\begin{eqnarray}
\frac{\partial I}{\partial \psi } &=&\frac{4\pi }{c}J_{\vartheta }, \\
\frac{\partial I}{\partial \vartheta } &=&\frac{4\pi }{c}J_{\psi },
\end{eqnarray}%
yielding the constraint%
\begin{equation}
\frac{\partial J_{\psi }}{\partial \psi }=\frac{\partial J_{\vartheta }}{%
\partial \vartheta }  \label{solubility}
\end{equation}%
which is a solubility condition for the structure functions. In
this regard we notice that as a consequence of the kinetic
constraints the function $I$ in the previous equations is of the
form $I(\psi ,\vartheta ,\varepsilon _{M}^{k}t)$, i.e., in
contrast to Paper I it is no longer a flux-function. Therefore,
the solubility condition (\ref{solubility}) can always be
satisfied. Eqs.(\ref{GSapp})-(\ref{solubility}) therefore provide
consistent solutions for both poloidal and toroidal self magnetic
fields in a collisionless AD plasma.

It is remarkable that in principle all of the populations of
charged particles (PPs, BPs and TPs) can contribute to the
generation of the toroidal magnetic field. More precisely, the
following mechanisms can be involved:

\#1) FLR and diamagnetic effects, driven by temperature anisotropy, of the
same kind as those described in Paper I;

\#2) Parallel velocity perturbations $U_{\parallel \ast s}^{\prime }$, which
generate a poloidal flow velocity, giving a related contribution to the
electric current density through $J_{\psi }$ and $J_{\vartheta }$;

\#3) FLR effects driven by the remaining thermodynamic forces (see
Section 5). These contributions are produced by the diamagnetic
KDF and arise because of the asymptotic ordering introduced here;

\#4) Gyrophase-dependent contributions driven by the same thermodynamic
forces. These are originated by the inverse GK transformation of the
guiding-center quantities in the QSA-KDF.

As discussed above, contributions \#2 and \#4 were negligible under the
circumstances discussed in Paper I. Therefore they should be considered as
characteristic features of open-field configurations.

We refer to the mechanism of self-generation of both poloidal and
toroidal magnetic fields as a \textit{quasi-stationary kinetic
dynamo effect}. In contrast to customary MHD treatments, this type
of dynamo effect occurs \textit{in the absence of possible
instabilities or turbulence phenomena}. In particular, in the case
of TPs, the self generation of toroidal field could take place
even \textit{without any net accretion} in the domain of interest,
in presence of open magnetic field lines. This phenomenon is
analogous to that treated in Paper I for closed-field
configurations. In particular, the toroidal field is associated
with the existence of torques which cause redistribution of
angular momentum, producing radial inflows and outflows of disk
material. As a consequence, various scenarios can be envisaged in
which stationary radial flows and kinetic dynamos are present in
AD plasmas, both affected by processes of type \#1-\#4.

\bigskip

\section{Quasi-stationary accretion flow}

Let us now consider specifically the application of the kinetic
solution developed here to the investigation of the accretion
process in AD plasmas.

The inward accretion flow in ADs is usually ``slow'' in comparison with the
characteristic Larmor time $\tau_{Ls}$. For example, AD plasmas with $%
B\sim10^{1}-10^{8}G$ have Hydrogen-ion Larmor rotation times in the range $%
\tau _{Li}\sim 10^{-4}-10^{-11}s$ which is shorter than the dynamical
timescale at most relevant radii. For typical plasma densities and
temperatures in the range $n_{i}\sim 10^{9}-10^{11}cm^{-3}$ and $T_{i}
\sim1-10keV$, the (Spitzer) ion collision time (below which the
plasma can be considered collisionless) is in the range $\tau _{C}\sim
10^{2}-10^{5}s$ (the upper value corresponding to high temperature and low
density). Independent of the physical origin of the accretion process, we
can therefore expect that the present theory correctly describes phenomena
occurring on all time-scales in the range $\tau _{Li}<t<\tau _{C}$.

We next determine the local poloidal and radial flow velocities for the
various particle sub-species. By definition, these are given by%
\begin{eqnarray}
&&\left. V_{ps}\equiv \mathbf{V}_{s}\cdot \mathbf{e}_{p}=\right.  \notag \\
&&\left. =\sum\limits_{sub-species}\frac{1}{n_{s}^{tot}}\int_{\Gamma
_{u}^{l}}d^{3}v\left[ \mathbf{v}\cdot \mathbf{e}_{p}\right] \widehat{f_{\ast
s}}\left[ 1+g_{s}^{\prime }\right] ,\right. \\
&&\left. V_{Rs}\equiv \mathbf{V}_{s}\cdot \mathbf{e}_{R}=\sum%
\limits_{sub-species}\frac{1}{n_{s}^{tot}}J_{Rs}^{l},\right. \\
&&\left. J_{Rs}^{l}\equiv \int_{\Gamma _{u}^{l}}d^{3}v\left[ \mathbf{v}\cdot
\mathbf{e}_{R}\right] \widehat{f_{\ast s}}\left[ 1+g_{s}^{\prime }\right]
,\right.
\end{eqnarray}%
where $\mathbf{e}_{p}\equiv \frac{\nabla \psi \times \nabla \varphi }{%
\left\vert \nabla \psi \times \nabla \varphi \right\vert }$ and $\mathbf{e}%
_{R}\equiv \frac{\nabla R}{\left\vert \nabla R\right\vert }$ and the
summations are performed over the particle sub-species for $l=T,B,P$. We
stress that the velocity-space integrals indicated above must contain the
contributions from PPs, BPs and TPs and so $%
J_{Rs}=J_{Rs}^{T}+J_{Rs}^{B}+J_{Rs}^{P}$, where $J_{Rs}^{T},$ $J_{Rs}^{B}$
and $J_{Rs}^{P}$ are the corresponding mass currents. As an example, let us
consider the leading-order contributions obtained ignoring FLR corrections.
Explicit calculation gives%
\begin{eqnarray}
V_{ps} &\cong &U_{\parallel s}\mathbf{b}\cdot \mathbf{e}_{p},  \label{Vp} \\
V_{Rs} &\cong &U_{\parallel s}\mathbf{b}\cdot \mathbf{e}_{R,}  \label{Vr}
\end{eqnarray}%
where $U_{\parallel s}\equiv U_{\parallel s}\left( \psi ,\vartheta
,\varepsilon _{M}^{k}t\right) \equiv \frac{I}{B}\xi _{s}$ and the
functional dependence $\xi _{s}=\xi _{s}\left( \psi ,\vartheta
,\varepsilon _{M}^{k}t\right) $ is prescribed by the kinetic
constraints (\ref{kinkin}). We stress that the precise form of
$\xi _{s}\left( \psi ,\vartheta ,\varepsilon _{M}^{k}t\right) $
still has to be chosen to satisfy the solubility constraints
imposed by Ampere's law (see the discussion in previous section)
and so the radial mass current density $J_{Rs}$ is generally a
function of the form $J_{Rs}\equiv J_{Rs}\left(
\mathbf{x},\varepsilon
_{M}^{k}t\right) =\overline{J_{Rs}}(\psi ,\vartheta ,\varepsilon _{M}^{k}t)$%
. We are interested in situations where there is a \textit{net
radial accretion flow} i.e. where the average radial mass current
$\left\langle
\left\langle J_{Rs}\right\rangle \right\rangle \equiv \frac{1}{z_{2}-z_{1}}%
\int_{z_{1}}^{z_{2}}J_{Rs}dz$ (with $z_{1}$ and $z_{2}$ being suitably
prescribed) is negative. There are local contributions to $\left\langle
\left\langle J_{Rs}\right\rangle \right\rangle $ from TPs, BPs and PPs, but
the overall accretion flow is mainly associated with PPs.

Let us show that such a solution exists. We seek particular Vlasov-Maxwell
equilibria which are globally quasi-neutral, in the sense of Eq.(\ref{qneu}%
). These equilibria are uniquely defined once $\Phi
(\mathbf{x},\varepsilon _{M}^{k}t)$, $\psi \left(
\mathbf{x},\varepsilon _{M}^{k}t\right) $, $I(\psi ,\vartheta
,\varepsilon _{M}^{k}t)$ and the structure functions are
prescribed. The latter, by definition, are arbitrary smooth real
functions of the specified variables as required by the kinetic
constraints. Notice that, if the quasi-neutrality condition is
valid, an analytical solution for the ES potential can be obtained
as shown in Paper I. Furthermore, we consider an example in which
by assumption

\begin{enumerate}
\item $U_{\parallel s}$ is non-vanishing and dominant with respect to FLR
effects, so that Eqs.(\ref{Vp}) and (\ref{Vr}) apply.

\item Particular solutions have a definite parity property with
respect to the spatial reflection $z\rightarrow -z.$ As a specific
case, the poloidal flux $\psi $ is assumed here to be
antisymmetric, i.e., $\psi \left( z\right) =-\psi \left( -z\right)
$, while both the toroidal and poloidal magnetic fields are
symmetric. As a consequence, the toroidal current density must be
antisymmetric. This can be realized only if a vertical electric
field is present (i.e., one in the $z$ direction), consistent with
the quasi-neutrality condition.
\end{enumerate}

If these assumptions are valid, Ampere's law demands that, to
leading
order, i.e., neglecting diamagnetic FLR effects,%
\begin{eqnarray}
\frac{\partial I}{\partial \psi } &\cong &\frac{4\pi }{c}\sum%
\limits_{s}Z_{s}en_{s}^{tot}U_{\parallel s}, \\
\frac{\partial I}{\partial \vartheta } &\cong &0,
\end{eqnarray}%
namely $I\cong I\left( \psi \right) $ at this order of
approximation. Therefore, in this case a solution consistent with
the requirement of net radial accretion flow and Vlasov-Maxwell
equilibrium is obtained imposing that the species number density
$n_{s}^{tot}$ is even in $\psi $, while the species structure
function $\xi _{s}$ is odd with respect to the same variable. A
solution of this type is consistent with the angular momentum
conservation law (\ref{angcons}); in order to obtain the solution,
suitable kinetic boundary conditions must be prescribed (see the
discussion following Eq.(\ref{dopo})). This proves that stationary
accretion solutions exist and are admitted by the present kinetic
theory for the \textquotedblleft
incoming\textquotedblright\ QSA-KDF, namely for $\left. \widehat{f_{\ast s}}%
\right\vert _{\vartheta _{2}}$ in the subset $\mathbf{v}\cdot \mathbf{e}%
_{R}<0$ (see Fig.2). The same conclusion is in principle applicable for
outflows, by appropriate prescription of the \textquotedblleft
outgoing\textquotedblright\ QSA-KDF $\left. \widehat{f_{\ast s}}\right\vert
_{\vartheta _{2}}$ in the subset $\mathbf{v}\cdot \mathbf{e}_{R}>0$. In
fact, the angular momentum conservation law (\ref{angcons}) allows both
inward and outward radial fluid velocities for each species, namely having $\mathbf{%
V}_{s}^{tot}\cdot \mathbf{e}_{R}<0$ or $>0$ respectively. Indeed,
for a collisionless plasma the species tensor pressure is
generally non isotropic (see the related discussions in Paper I
and Ref.\cite{Catania2}) such that Eq.(\ref{angcons}) is
identically satisfied. Unlike the customary view based on ideal
MHD, for which a self-consistent treatment of inflow and outflow
solutions is usually difficult, within the present theory both
inflows and outflows can occur independently and are described
consistently by their respective QSA-KDFs. In particular,
Eq.(\ref{angcons}) shows that radial flows arise due both to the
parallel velocities $U_{\parallel s}$ and to the kinetic effects
carried by the FLR diamagnetic corrections. As a result, species
radial flow velocities appear necessarily in combination with
non-isotropic tensor pressures and a non-vanishing toroidal
magnetic field. In conclusion, the theory predicts the possibility
of having purely inflowing matter in quasi-stationary AD plasmas,
or of having co-existing inflows and outflows.

\bigskip

\bigskip

\section{Conclusions}

In this paper, a consistent theoretical investigation of the slow kinetic
dynamics of collisionless non-relativistic and axisymmetric AD plasmas has
been presented. The formulation is based on a kinetic approach developed
within the framework of the Vlasov-Maxwell description. We have considered
here plasmas immersed in quasi-stationary magnetic fields characterized by
open nested magnetic surfaces. This can be appropriate for radiatively
inefficient accretion flows onto black holes, some of which are believed to
be associated with a plasma of collisionless ions and electrons having
different temperatures, and there can be other related applications to the
inner regions of accretion flows onto magnetized neutron stars and white
dwarfs. The discussion presented here provides a background for future
investigations of instabilities and turbulence occurring in these plasmas.

We have shown that a new type of asymptotic kinetic equilibria exists, which
can be described by QSA-KDFs expressed in terms of generalized bi-Maxwellian
distributions. These solutions permit the consistent treatment of a number
of physical properties characteristic of collisionless plasmas. The
existence of these equilibrium solutions has been shown to be warranted by
imposing suitable kinetic constraints for the structure functions entering
the definition of the QSA-KDFs. In terms of these solutions, the slow
dynamics of collisionless AD plasmas has been described by means of a
suitable reduced GK-Vlasov equation. In addition, the theory permits the
consistent treatment of gravitational EM particle trapping phenomena,
allowing one to distinguish between different populations of charged
particles.

We have shown that the kinetic approach is suitable for the description of
quasi-stationary AD plasmas subject to accretion flows and kinetic dynamo
effects responsible for the self-generation of both poloidal and toroidal
magnetic fields. Four intrinsically-kinetic physical mechanisms have been
included in the treatment of this, related to temperature anisotropy,
parallel velocity perturbations and FLR-diamagnetic effects.

The novelty of the present approach, with respect to traditional fluid
treatments, lies in the possibility of explicitly constructing asymptotic
solutions for the fluid equations: the calculation of all of the relevant
fluid fields involved (e.g. the plasma charge and mass current densities and
the radial flow velocity) can be performed in a straightforward way using a
species-dependent asymptotic expansion of the QSA-KDF.

We believe that this study makes a relevant contribution for the description
of two-temperature collisionless AD plasmas and the improvement of our
understanding of their physical properties. The kinetic treatment developed
here can also provide a convenient starting point for making a kinetic
stability analysis of these plasmas.

\bigskip

\section{Acknowledgments}

This work has been partly developed in the framework of MIUR (Italian
Ministry of University and Research) PRIN Research Programs and the
Consortium for Magnetofluid Dynamics, Trieste, Italy.

\end{document}